\documentclass[a4paper,10pt]{iopart}
\pdfoutput=1
\usepackage[utf8]{inputenc}
\usepackage{graphicx}
\usepackage{siunitx}
\usepackage{booktabs}
\usepackage{url}
\expandafter\let\csname equation*\endcsname\relax
\expandafter\let\csname endequation*\endcsname\relax
\usepackage[version=4]{mhchem}
\usepackage{cite}
\usepackage{amsmath,amssymb}
\usepackage{color}
\begin{document}
\title[Electron swarm parameters in \ce{C2H2}, \ce{C2H4} and \ce{C2H6}]
{Electron swarm parameters in \ce{C2H2}, \ce{C2H4} and \ce{C2H6}: 
measurements and kinetic calculations}
\author{N R Pinh\~ao$^1$, D Loffhagen$^2$, M Vass$^3$, P Hartmann$^3$, 
I Korolov$^4$, S Dujko$^5$, D Bo\v{s}njakovi\'{c}$^5$ and Z Donk\'o$^3$}

\address{$^1$ Instituto de Plasmas e Fus\~ao Nuclear, Instituto Superior T\'ecnico, 
Universidade de Lisboa, Av. Rovisco Pais, 1049-001 Lisboa, Portugal}
\address{$^2$ Leibniz Institute for Plasma Science and Technology, 
Felix-Hausdorff-Str.~2, 17489 Greifswald, Germany}
\address{$^3$ Institute for Solid State Physics and Optics, Wigner Research 
Centre for Physics, 1121 Budapest, Konkoly Thege Mikl\'os str. 29-33, Hungary}
\address{$^4$ Department of Electrical Engineering and Information Science, 
Ruhr-University Bochum, D-44780, Bochum, Germany}
\address{$^5$ Institute of Physics, University of Belgrade, Pregrevica 118, 
11080 Belgrade, Serbia}
\ead{npinhao@ctn.tecnico.ulisboa.pt}

\begin{abstract}
This work presents swarm parameters of electrons (the bulk drift velocity, the 
bulk longitudinal component of the diffusion tensor, and the effective ionization frequency) 
in \ce{C2H_n}, with $n =$ \numlist{2;4;6}, measured in a scanning drift tube 
apparatus under time-of-flight conditions over a wide range of the reduced 
electric field, $\SI{1}{Td}\leq E/N\leq \SI{1790}{Td}$ (\SI{1}{Td} = \SI{e-21}{Vm^2}). 
The effective steady-state Townsend ionization coefficient is also derived from the experimental
data. A kinetic simulation of the experimental drift cell allows estimating the 
uncertainties introduced in the data acquisition procedure and provides a 
correction factor to each of the measured swarm parameters. 
These parameters are compared to results of previous experimental studies, as  well 
as to results of various kinetic swarm calculations: solutions of the electron 
Boltzmann equation under different approximations (multiterm and density gradient
expansions) and Monte Carlo simulations. The experimental data are consistent with 
most of the swarm parameters obtained in earlier studies.  In the case of 
\ce{C2H2}, the 
swarm calculations show that the thermally excited vibrational 
population should not be neglected, in particular, in the fitting of cross 
sections to swarm results. 
\end{abstract}


\submitto{\PSST}
\maketitle

\section{Introduction}

\label{sec:intro}

Acetylene (\ce{C2H2}), ethylene (\ce{C2H4}) and ethane (\ce{C2H6}) are relatively 
simple hydrocarbons useful in specialized applications such as plasma-assisted 
combustion~\cite{Adamovich:2014,Starikovskiy:2013,kosarev2009,kosarev2013, 
kosarev2015,kosarev2016}, the fabrication of diamond-like films~\cite{Robertson:2002},
graphene and carbon nanostructures~\cite{Kumar:2010}, and particle 
detectors~\cite{Fonte:2010}. They are also involved in various chemical reactions 
in fusion plasmas~\cite{VonKeudell:2001}, the Earth's atmosphere~\cite{Varanasi:1983} 
and in planetary atmospheric chemistry~\cite{Courtin:1984}.

Knowledge on both electron collision cross sections and electron swarm parameters 
is needed for the quantitative modelling of plasmas. However, with the exception 
of the drift velocity, which was measured e.g.\ in~\cite{Hasegawa_Date:2015,Nakamura:2010, Cottrell:1968,Bowman:1967,Cottrell:1965} for \ce{C2H2}, in~\cite{Hasegawa_Date:2015, Takatou:2011,Schmidt:1992,Bowman:1967,Wagner:1967,Christophorou:1966,Cottrell:1965, 
Hurst:1963,Bortner:1957} for \ce{C2H4}, and in~\cite{Hasegawa_Date:2015,Shishikura:1997, Kersten:1994,Schmidt:1992,Cottrell:1968,Bowman:1967,Cottrell:1965} for \ce{C2H6}, 
further experimental transport and ionization coefficients have less frequently 
been reported for these hydrocarbon gases. 
Measurements of the longitudinal component of the diffusion tensor under time-of-flight (TOF) 
conditions were additionally reported in~\cite{Nakamura:2010} for~\ce{C2H2}, 
\cite{Takatou:2011,Schmidt:1992,Wagner:1967} for \ce{C2H4}, 
and~\cite{Shishikura:1997,Schmidt:1992} for \ce{C2H6}. 
Hasegawa and Date~\cite{Hasegawa_Date:2015} also determined the  effective ionization 
coefficient by the steady-state Townsend (SST) method for seven organic gases 
including acetylene, ethylene, and ethane. In addition to the drift velocity for
\ce{C2H6}, Kersten~\cite{Kersten:1994} measured the effective ionization
coefficient under TOF conditions for a narrow range of the reduced  electric field, $E/N$. 
Furthermore, measured data for the effective SST ionization coefficient have been 
reported e.g.\ in~\cite{Heylen:1963} for \ce{C2H2}, in~\cite{Heylen:1978,Heylen:1963}
for \ce{C2H4}, and in~\cite{Watts:1979,Heylen:1975,LeBlanc:1960} for \ce{C2H6}.

The aim of this work is (i) to determine the electron transport and ionization 
coefficients in \ce{C2H2}, \ce{C2H4} and \ce{C2H6} gases in a wide range of $E/N$, 
(ii) to compare these results with those obtained in earlier investigations 
of these gases, and (iii) to compare the experimental data with those obtained from
kinetic calculations and simulations using up-to-date electron collision cross section sets. 

\begin{figure}[tp]
 \centering
 \includegraphics[width=0.5\linewidth]{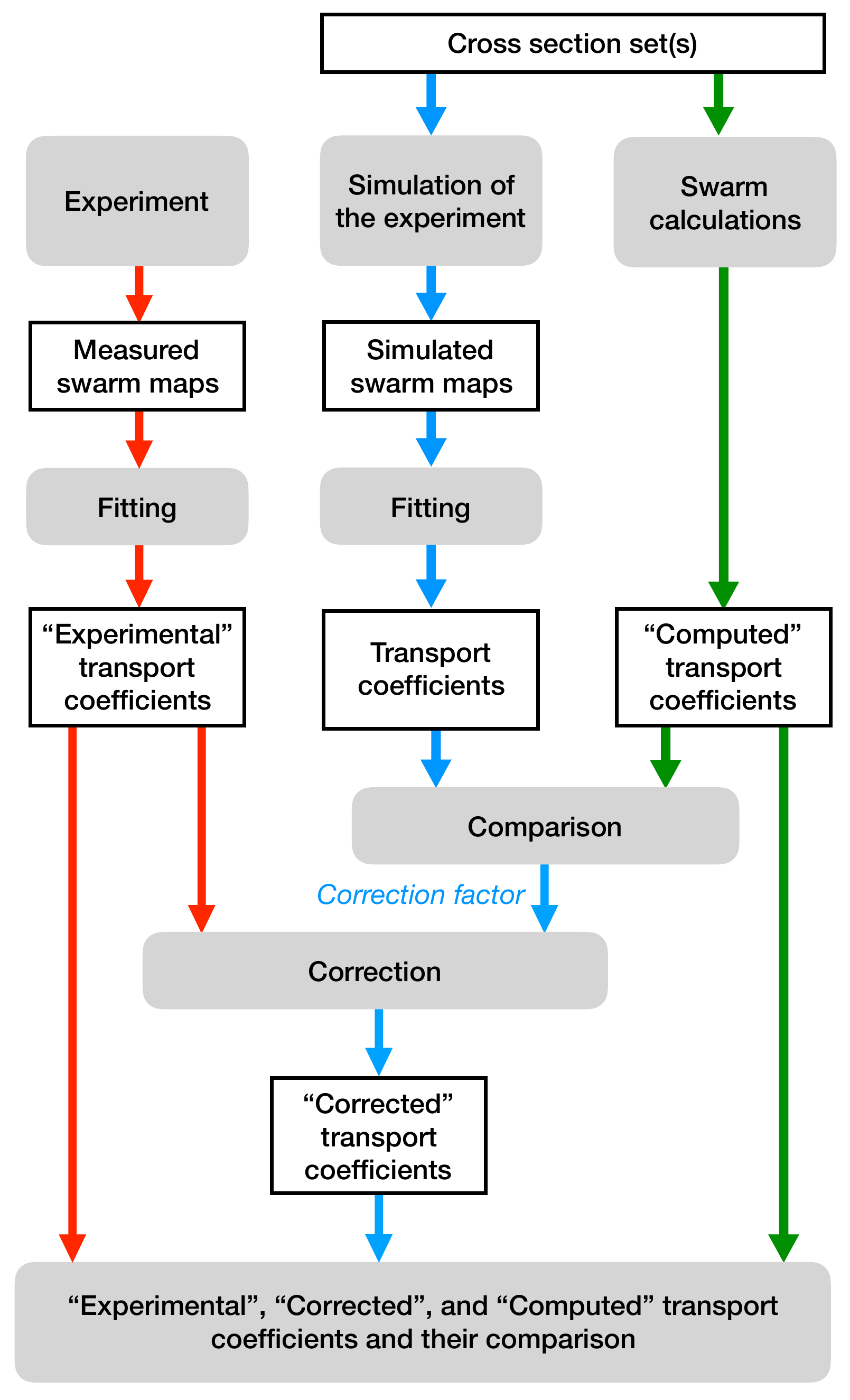}
 \caption{Graphical representation of the work reported in this article. The red 
 arrows indicate the path from the measurements to the 'Experimental' transport
 coefficients and ionization frequencies via fitting of the measured 'swarm maps'. 
 Another 'Corrected' set 
 of experimental data is also derived based on a correction procedure which is 
 aided by simulations of the experimental setup and related data acquisition 
 (indicated by blue arrows) and by kinetic computations of the swarm parameters.   
 The results of these calculations ('Computed' transport coefficients) 
 are also compared to the experimental data (green arrows).} \label{fig:flow}
\end{figure}

The workflow of our studies can be followed with the aid of figure~\ref{fig:flow}. 
The red arrows show the path to the 'Experimental transport coefficients' 
including the  effective ionization frequencies. The first step along this path 
consists of the measurements carried out with our scanning drift 
tube apparatus. This is a pulsed system, which is described in section~\ref{Sec Exp}. 
It records current traces generated by electrons collected from clouds that arrive 
after having flown over the drift region. 
The results of the experiments are the so-called 'swarm maps' which are collections 
of these current traces for a number of drift gap length values. The swarm parameters 
are derived from the measured swarm maps via a fitting procedure that 
assumes that the  current measured in the experiment is proportional to the electron 
density. For the fitting we use the theoretical form of the electron density in the 
presence of an electric field pointing in the $-z$ direction and under TOF conditions:
\begin{equation}
n_{\rm e}(z,t) = \frac{n_0}{(4 \pi D_{\rm L} t)^{1/2}} \exp 
\biggl[ \nu_{\rm eff} t - \frac{(z - Wt)^2}{4 D_{\rm L} t} \biggr].
\label{eq:n}
\end{equation}
This is the solution of the spatially one-dimensional (1D) continuity equation and 
represents a Gaussian pulse drifting along the $z$ direction with the bulk drift 
velocity, $W$, and diffuses along the centre-of-mass according to the bulk 
longitudinal component of the diffusion tensor $D_{\rm L}$. Here $n_0$ is the electron density 
at $z=0$ at time $t=0$, and $\nu_{\rm eff}$ is the effective ionization frequency. 
From the fitting procedure we obtain $W$, $D_\mathrm{L}$, and $\nu_\mathrm{eff}$. 
The application of the relation~\cite{Blevin:1984}
\begin{equation}
\frac{1}{\alpha_\mathrm{eff}} = \frac{W}{2 \nu_{\rm eff}} + 
\sqrt{\left( \frac{W}{2 \nu_{\rm eff}} \right)^2 - \frac{D_{\rm L}}{\nu_{\rm eff}}}
\label{eq:Blv1}
\end{equation}
allows us to derive the effective SST ionization coefficient, $\alpha_\mathrm{eff}$, 
as well.

The assumption that the measured current is proportional to the electron density 
is, in fact, an approximation, due to two reasons. First, the measured current is 
generated by moving charges in the detector of the system (see later). In our 
previous work \cite{Donko:2019} we have found that the detection sensitivity depends 
on the gas pressure and the collision cross sections, which both influence the free 
path of the electrons. This means that any variation of the energy distribution along 
the $z$ direction in the electron cloud may results in a distortion of the detected 
pulse and a deviation from the analytical fitting function (\ref{eq:n}) assumed. 
Second, the measured current is proportional to the electron flux consisting of the 
advective and diffusive component. The advective component is proportional to the 
electron density, where the coefficient of proportionality is the flux drift velocity, 
and the diffusive component is proportional to the gradient of the electron density. 
Using Ramo's theorem~\cite{R}, it can be shown that for the experimental conditions 
considered in the present work, the contribution of the diffusive component to the 
current is negligible compared to the contribution of the advective component, except 
in the early stage of the swarm development when the spatial gradients of the 
electron density are more significant. 
 
The errors introduced by the first effect mentioned above can be quantified by a 
procedure, which is marked by blue arrows in figure~\ref{fig:flow}. We carry out 
a (Monte Carlo) simulation of the electrons' motion in the experimental system. 
This simulation generates the \textit{same} type of swarm maps, which are obtained 
in the experiments, and a set of swarm parameters is derived via the 
\textit{same} fitting procedure as in the case of experimental swarm maps. The 
transport coefficients and ionization frequencies obtained in this way are compared 
with the 'Computed' ones, originating from kinetic swarm calculations. We note that 
(i) this comparison does not include any experimental data, (ii) the system's simulations 
use the \textit{same} cross section set as in the kinetic swarm calculations, and 
(iii) uncertainties of the used collision cross sections have little influence on 
the outcome of the comparison of the parameter sets obtained by swarm 
calculations and simulations of the experimental system. The result of this comparison 
is gas- and $E/N$-dependent correction factors that are applied to the experimental 
data, providing sets of 'Corrected' experimental transport and ionization coefficients. 
Details are given below in section~\ref{sec:correction}. 

The two (raw and corrected) sets of experimental results are compared with swarm 
parameters derived from kinetic calculations based on solutions of the electron Boltzmann
equation and on Monte Carlo simulations as described in detail in section~\ref{Sec 3}. 
The application of these different approaches allows us to mutually verify the 
accuracy of the different methods and test the assumptions used by each method. 
The 'flow' of this process is indicated by the green arrows in figure~\ref{fig:flow}.

The manuscript is organized as follows: in section \ref{Sec Exp} we give a concise 
description of our experimental setup. A discussion of the various computational 
methods and the resulting swarm parameters is presented in section~\ref{Sec 3}, 
and section~\ref{sec:correction} describes the correction procedure applied to the 
experimental data. It is followed by the discussion of the results in section 
\ref{Sec:results}. This section comprises the presentation of the present 
experimental results for each gas and their comparison with previously available 
measured data as well as the comparison between transport parameters and ionization 
coefficients computed using the various numerical methods and the present experimental 
data. Section \ref{Sec 6} gives our concluding remarks and in the appendix we provide 
tabulated values of our experimental results.

\section{Experimental system}
\label{Sec Exp}

\begin{figure}[tp]
 \centering
\includegraphics[width=0.8\linewidth]{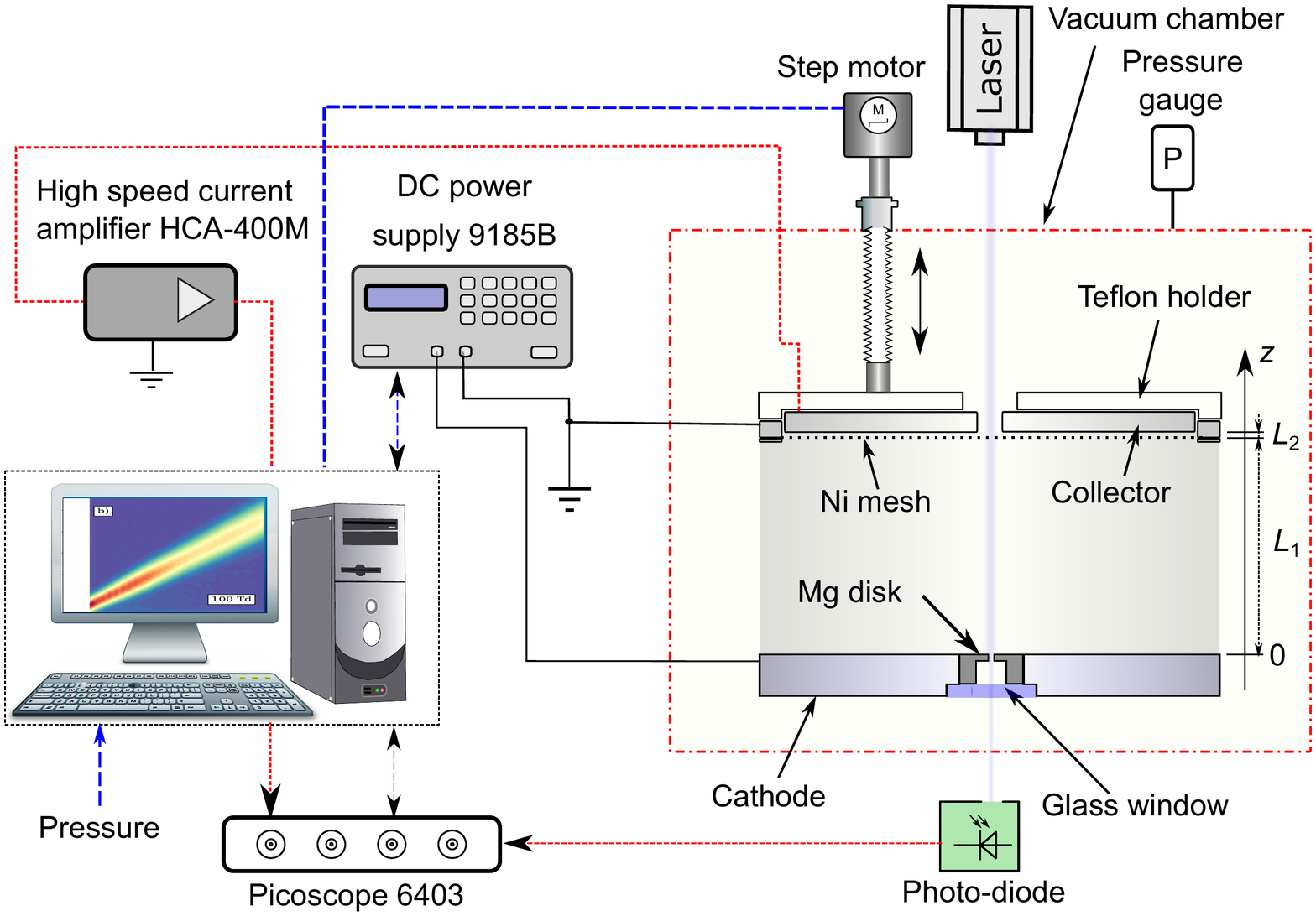}
 \caption{Simplified schematic of the scanning drift tube apparatus}
 \label{fig:exp}
\end{figure}

The experiments are based on a 'scanning' drift tube apparatus, of which the details 
have been presented in~\cite{Korolov:2016}. This apparatus has already been applied 
for the measurements of transport and ionization coefficients of electrons in various 
gases: argon, synthetic air, methane, deuterium~\cite{Korolov_2016} and carbon 
dioxide~\cite{Vass:2017}. 
In contrast to previously developed drift tubes, our system allows for recording 
of 'swarm maps' that show the spatio-temporal development of electron clouds under 
TOF conditions. The simplified scheme of our experimental apparatus is shown in 
figure~\ref{fig:exp}. 

The drift cell is situated within a vacuum chamber made of stainless steel. The 
chamber can be evacuated by a turbomolecular pump backed with a rotary pump to a 
level of about \SI{1e-7}{mbar}. The pressure of the working gases inside the 
chamber is measured by a Pfeiffer CMR 362 capacitive gauge. 

Ultraviolet light pulses (\SI{1.7}{\micro\joule}, \SI{5}{ns}) of a frequency-quadrupled 
diode-pumped YAG laser enter the chamber via a feedthrough with a quartz window 
and fall on the surface of a Mg disk used as photoemitter. This disk is placed at 
the centre of a stainless steel electrode with 105\,mm diameter that serves as the 
cathode of the drift cell. The detector that faces the cathode at a distance $L_1$ 
consists of a grounded nickel mesh (with $T = \SI{88}{\percent}$ 'geometric' 
transmission and 45 lines/inch density) and a stainless steel collector electrode 
that is situated at a distance of $L_2 = \SI{1}{\milli\metre}$ behind the mesh. 

Electrons emitted from the Mg disk fly towards the collector under the influence of 
an accelerating voltage that is applied to the cathode. This voltage is established 
by a BK Precision 9185B power supply. Its value is adjusted according to the required 
$E/N$ for the given experiment and the actual value of the gap ($L_1$) during the 
scanning process, where $E/N$ is ensured to be fixed. The current of the detector 
system is generated by the moving charges within the mesh-collector gap: according 
to the Shockley-Ramo theorem \cite{R,S,SH} the current induced by an electron moving 
in a gap between two plane-parallel electrodes with a velocity $v_z$ perpendicular 
to the electrodes is $I = - e_0 v_z / L$, where $-e_0$ is the charge of the electron 
and $L$ is the distance between the electrodes ($L=L_2$ in our case). Accordingly, 
in our setting the measured current at a given time $t$ is
\begin{equation}
I(t) = c \sum_{k} v_{z,k}(t) \, ,
\label{eq:ramo}
\end{equation}
where $c$ is a constant. The summation goes over all electrons being present in 
the volume bounded by the mesh and the collector at time $t$, and $v_{z,k}$ 
is the velocity component of the $k$-th electron in $z$ direction. 

Electrons entering the detector region (the gap between the nickel mesh and the 
collector) contribute to the measured current until their first collisions with 
the gas molecules, as these collisions randomise the angular distribution of their 
velocities. Therefore, the free path of the electrons plays a central role in the 
magnitude of the current. For conditions when this free path is longer than the 
detector gap, the electron sticking property of the collector surface plays a 
crucial role too, as reflected electrons generate a current contribution with an 
opposite sign with respect to that generated by the 'incoming' electrons. 
According to the above effects, which have been explored to some details 
in~\cite{Donko:2019}, the sensitivity of the detector changes with the nature of 
the gas (magnitudes and energy dependence of the electron collision cross sections), 
the pressure, as well as the energy distribution of the electrons. This dependence 
is the primary reason which calls for a correction of the measured transport and 
ionization coefficients as discussed in more details in section \ref{sec:correction}.

The collector current is amplified by a high speed current amplifier (type Femto 
HCA-400M) connected to the collector, with a virtually grounded input, and is 
recorded by a digital oscilloscope (type Picoscope 6403B) with sub-ns time 
resolution. Data collection is triggered by a photodiode that senses the laser 
light pulses. The low light pulse energy necessitates averaging over typically
\numrange{20000}{150000} pulses. The experiment is fully controlled by a 
computer using LabView software. 

\begin{figure}[tp]
 \centering
\includegraphics[width=0.4\linewidth]{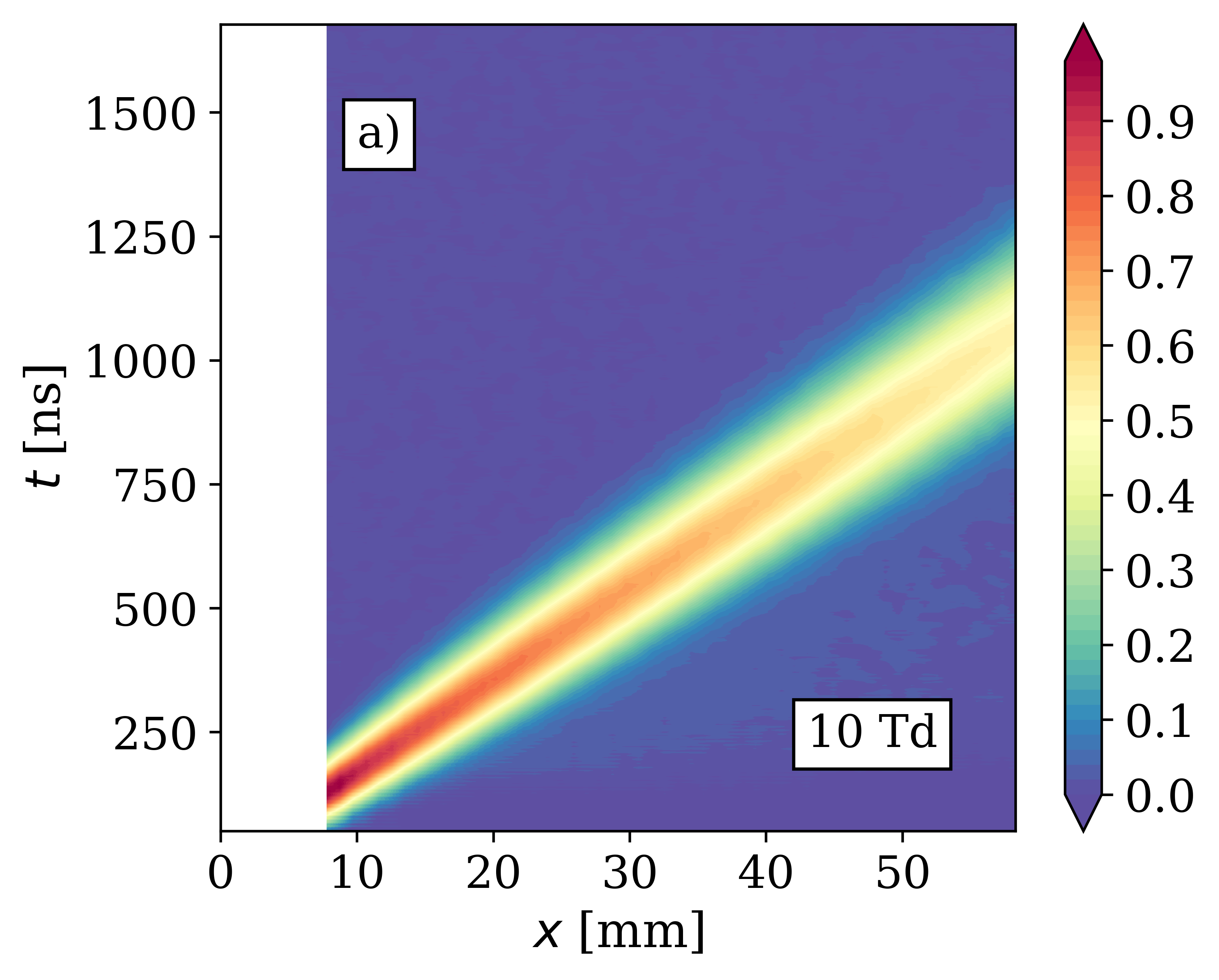}~~~
\includegraphics[width=0.4\linewidth]{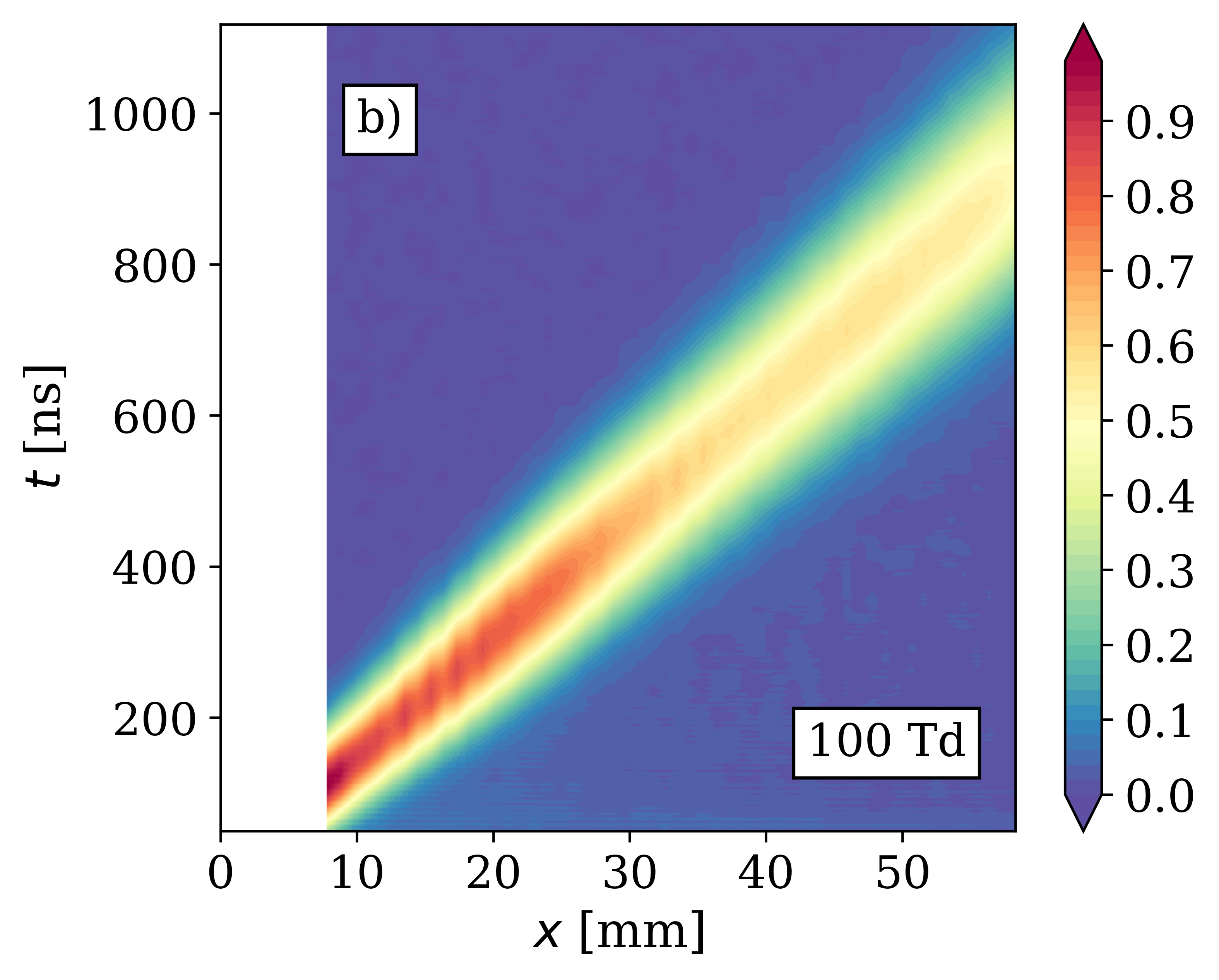}\\
\includegraphics[width=0.4\linewidth]{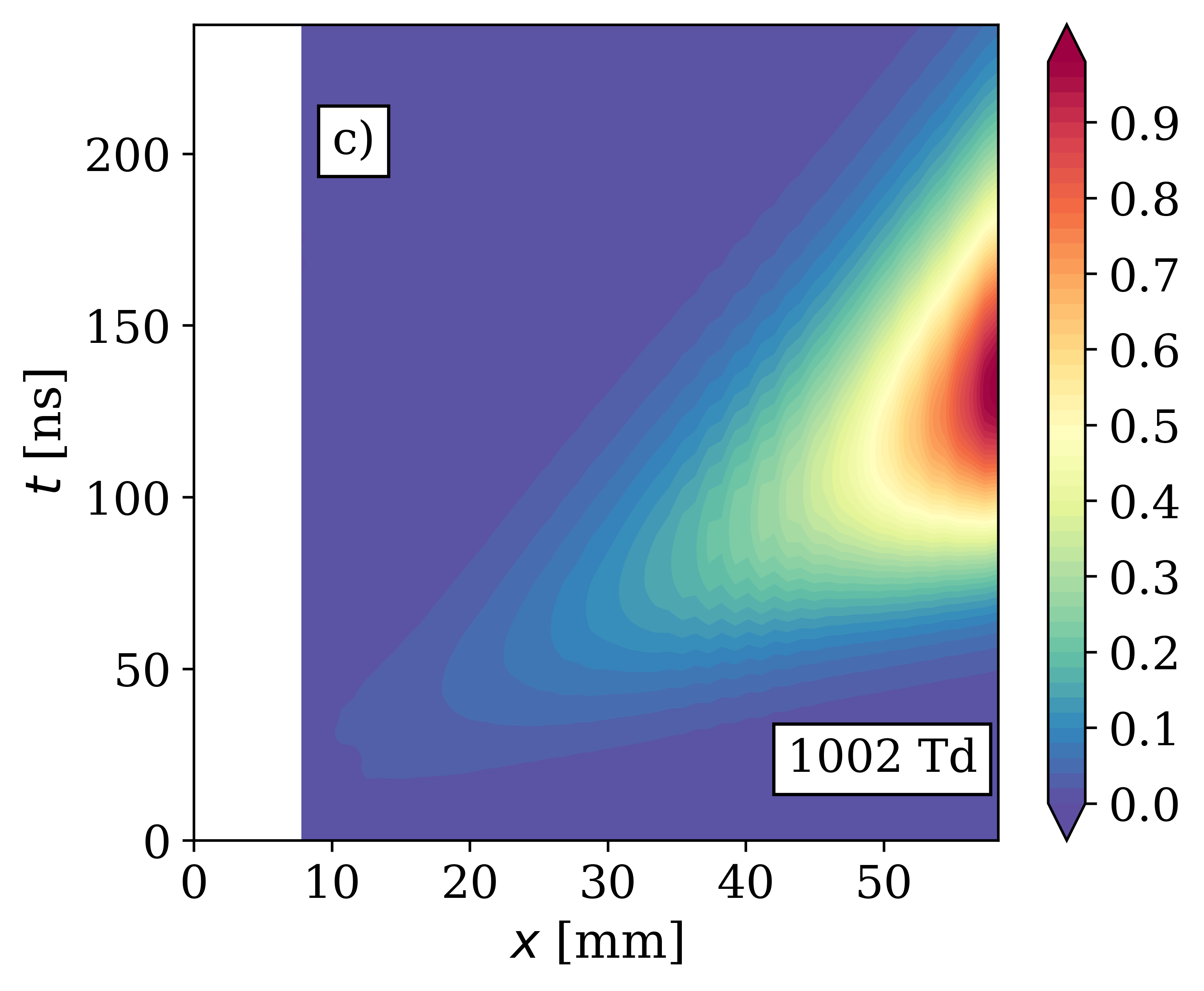}~~~
\includegraphics[width=0.4\linewidth]{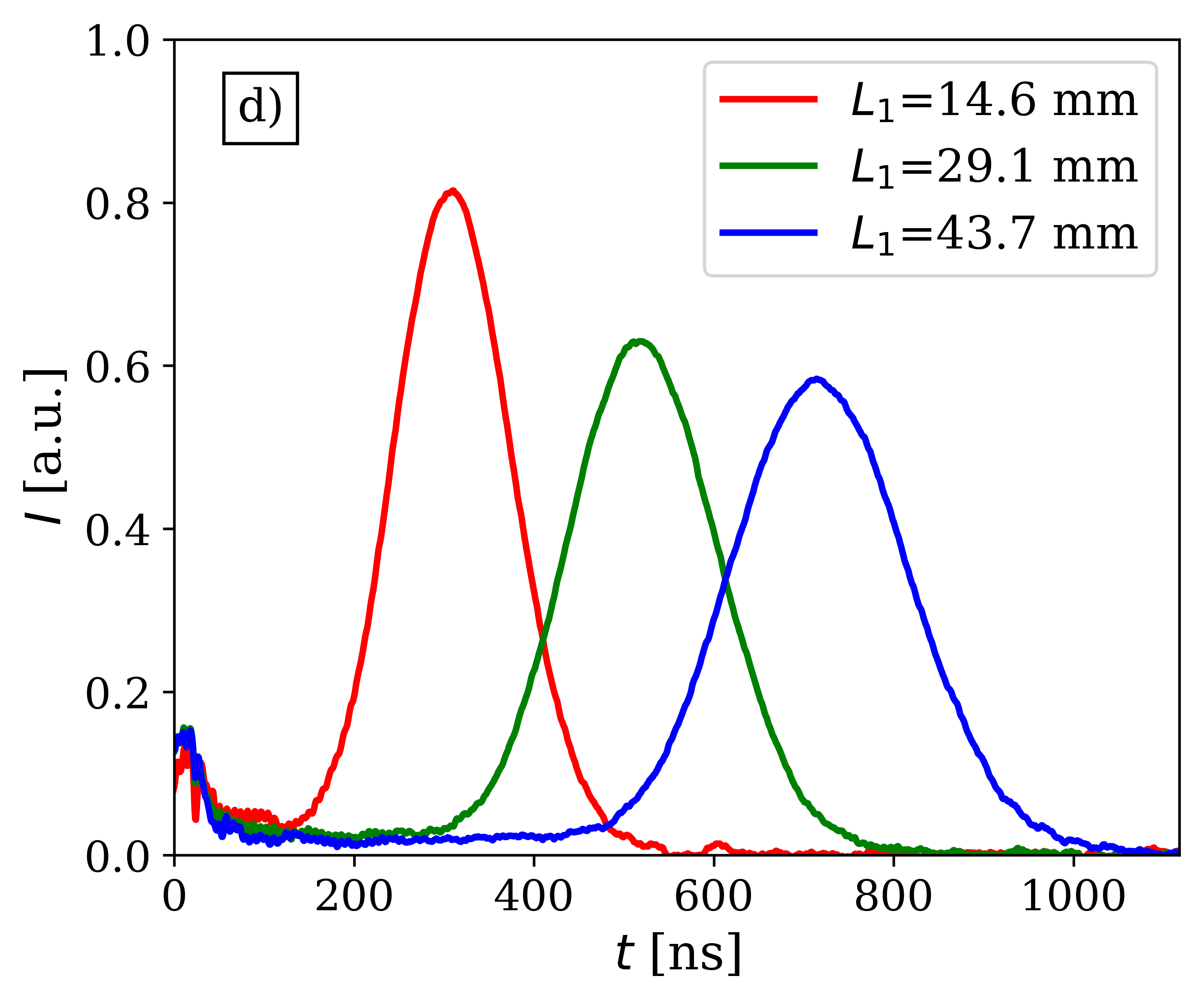}\\
 \caption{(a-c) Swarm maps recorded in \ce{C2H6} for different values of $E/N$, 
 as indicated. (d) Vertical cuts of the swarm map of (b), which are the measured 
 current traces at the drift length values given in the legend. The pulses have 
 nearly Gaussian shapes. The 'shift' of the pulses with increasing drift length 
 ($L_1$) is the manifestation of the drift, while their  widening is due to 
 (longitudinal) diffusion. As ionization in \ce{C2H6} is weak at $E/N$ = \SI{100}{Td}, 
 the amplitude of the pulses decreases with increasing $L_1$ due to the widening 
 of the pulse.}
 \label{fig:maps}
\end{figure}

\sisetup{range-units = brackets}%
During the course of the measurements current traces are recorded for several values 
of the gap length. The grid and the collector are moved together by a step motor 
connected to a micrometer screw mounted via a vacuum feedthrough to the vacuum 
chamber. The distance between the cathode and the mesh, i.e. the 'drift length', 
can be set within a range of $L_1$ = \SIrange{7.8}{58.3}{mm}. Here, we use 53 
positions within this domain.

Sequences of the measured current traces are subsequently merged to form 'swarm 
maps', which provide information about the spatio-temporal development of the 
electron cloud. Figure~\ref{fig:maps}(a)-(c) illustrates such swarm maps, obtained 
in experiments on \ce{C2H6}, for different values of the reduced electric field. 
The qualitative behaviour of the swarm is directly seen in these pictures: the 
slope of the region with appreciable current indicates the drift of the cloud, 
the widening of this region is related to (longitudinal) diffusion, while an 
increasing amplitude (as seen in panel (c)) is an indication of ionization. 
Figure~\ref{fig:maps}(d) displays vertical 'cuts' of the map shown in panel (b), 
for $E/N$ = \SI{100}{Td}. These cuts are, actually, the current traces recorded 
in the measurements at different gap length values. 

\section{Simulation of the electron swarm}
\label{Sec 3}

The experimental studies of the electron transport are supplemented by numerical 
modelling and simulation. In addition to Monte Carlo (MC) simulations, three 
different methods are applied to solve the Boltzmann equation (BE) for electron 
swarms in a background gas with density $N$ and acted upon by a constant electric 
field, $\vec{E}$\/: a multiterm method for the solution of the time-independent 
Boltzmann equation under spatially homogeneous and SST conditions, respectively, 
and the $S_n$ method applied to a density gradient expansion of the electron 
velocity distribution function (EVDF). They differ in their initial physical 
assumptions and in the numerical algorithms used and provide different properties 
of the electron swarms. 

Details of the different Boltzmann equation methods, as well as main aspects of 
the MC simulation have been discussed in~\cite{Vass:2017}, and we just provide a brief 
discussion below.

In the following, the electric field is parallel to the $z$ axis and points in the 
negative direction, $\vec{E} = -E\vec{e}_z$, and $\theta$ is the angle between 
$\vec{v}$ and $\vec{E}$. Moreover, we assume that the spatial and time scales, 
respectively, exceed the energy relaxation length and time, such that the transport 
properties of the electrons do not change with time $t$ and distance $z$ any longer. 
That is, the electrons have reached a hydrodynamic regime characterizing a state of 
equilibrium of the system where the effects of collisions and forces are dominant 
and the EVDF, $f(\vec{r},\vec{v},t)$, has lost any memory of the initial state.

We base our studies on the electron collision cross section sets from Song 
\textit{et al.}~\cite{Song:2017} for acetylene, Fresnet \textit{et al.}~\cite{Fresnet:2002} 
for ethylene and Shishikura \textit{et al.}~\cite{Shishikura:1997} for ethane. 
The cross sections for acetylene and ethane were extended to electron kinetic 
energies, $\epsilon$, of \SI{1000}{eV} by fitting a function with a 
$\log(\epsilon)/\epsilon$ dependence, according to the Born-Bethe high-energy 
approximation, to the tail of the original cross sections.  

The \ce{C2H2} data set includes the momentum transfer cross section for elastic 
collisions, three vibrational cross sections for single quanta excitation of modes 
$v_1/v_3$, $v_4/v_5$ and $v_2$ (the first two unresolved) and one vibrational cross 
section for two quanta excitation of $v_4+v_5$, three electronic excitation cross 
sections, the total electron-impact ionization cross section and the total dissociative 
electron attachment cross section for \ce{C2H2} leading to the formation of \ce{C2H-}, 
\ce{H-} and \ce{C2-}, respectively.

The \ce{C2H4} data set includes the momentum transfer cross section, two lumped 
vibrational cross sections with thresholds at \SIlist{0.118;0.365}{eV}, three 
electronic excitation cross sections, the total electron ionization cross section 
and a collision cross section for electron attachment.

Finally, the \ce{C2H6} set of collision cross sections includes the momentum 
transfer cross section, three lumped vibrational cross sections with thresholds 
at \SIlist{0.112;0.167;0.36}{eV}, two electronic excitation cross sections, the 
total electron ionization cross section and an electron attachment cross section.

All of the above cross section sets were developed neglecting the population of 
thermally excited vibrational states and superelastic processes. The implications 
of this approximation are discussed in section \ref{Sec: VibEx}.

\subsection{Boltzmann equation methods}

\subsubsection{Multiterm method for spatially homogeneous conditions}\label{Sec: BE0D}

In this approach, to describe $f(\vec{r}, \vec{v}, t)$ (abbreviated by BE 0D in the 
figures shown in section \ref{Sec:results}), we consider that the EVDF is spatially 
homogeneous (0D) and the electron density changes exponentially in time according 
to $n_\mathrm{e}(t) \propto \exp(\nu_\mathrm{eff}t)$ at the scale of the swarm. 
Here, the effective ionization frequency $\nu_\mathrm{eff}=\nu_{\rm i}-\nu_{\rm a}$ 
is the difference of the ionization ($\nu_{\rm i}$) and attachment ($\nu_{\rm a}$) 
frequencies. In this case we can neglect the dependence of $f$ on the space 
coordinates and write the EVDF under hydrodynamic conditions as
\begin{equation}
f(\vec{v},t) = \hat{F}(\vec{v})n_\mathrm{e}(t).
\end{equation}
The corresponding microscopic and macroscopic properties of the electrons are 
determined by the time-independent, spatially homogeneous Boltzmann equation for
$\hat{F}(\vec{v})$. As this distribution is symmetric around the field direction, 
it can be expanded with respect to the angle $\theta$ in Legendre polynomials 
$P_n(\cos\theta)$ with $n\geq0$. Substituting this expansion in the Boltzmann equation 
leads to a hierarchy of partial differential equations for the coefficients 
$\hat{f}_n(v)$ of this expansion. 
The resulting set of equations with typically eight expansion coefficients is solved 
employing a generalized version of the multiterm solution technique for weakly 
ionized steady-state plasmas~\cite{Leyh:1998} adapted to take into account the 
ionizing and attaching electron collision processes.

Using the computed expansion coefficients $\hat{f}_n(v)$, we obtain the effective 
ionization frequency, $\nu_\mathrm{eff}$, and the \textit{flux} drift velocity
\begin{equation}
w = -\mu E\,,\label{Eq:vd}
\end{equation}
where $\mu$ is the \textit{flux} mobility.
Explicit formulas of these transport parameters obtained by the BE 0D method can 
be found in~\cite{Vass:2017}. 

\subsubsection{Multiterm method for SST conditions}\label{Sec: BESST}

This approach to describe the EVDF (abbreviated by BE~SST in the figures shown 
in section \ref{Sec:results}) takes into account that $f(\vec{r},\vec{v},t)$ has 
reached SST conditions so that the mean transport properties of the electrons are 
time-independent, do not vary with position any longer, and the electron density
assumes an exponential dependence on the distance according to 
$n_\mathrm{e}(z) \propto \exp(\alpha_\mathrm{eff}z)$. 
Thus, we can neglect the dependence of $f$ on time and write the EVDF under SST 
conditions as
\begin{equation}
f(z,\vec{v}) = \tilde{F}^{(\rm S)}(\vec{v})n_\mathrm{e}(z), 
\end{equation}
where the upper index (S) denotes SST conditions. In accordance with the 
procedure described in section~\ref{Sec: BE0D}, the corresponding microscopic 
and macroscopic properties of the electrons are determined by the steady-state, 
spatially homogeneous Boltzmann equation for $\tilde{F}^{(\rm S)}(\vec{v})$. Since 
this distribution is again symmetric around the  direction of the field, it can 
be expanded in Legendre polynomials $P_n(\cos\theta)$ with $n \ge0$. 
The substitution of this expansion into the Boltzmann equation leads to a set of 
partial differential equations for the expansion coefficients 
$\tilde{f}^{(\rm S)}_n(v)$, which is solved efficiently by a modified version of the 
multiterm method~\cite{Leyh:1998} adapted to treat SST conditions~\cite{Vass:2017}. 

In this approach, the effective SST ionization coefficient is directly given by
\begin{equation}
\alpha_\mathrm{eff}=\frac{\nu_\mathrm{eff}^{(\rm S)}}{v_\mathrm{m}^{(\rm S)}} .
\label{eq:alphaeffBESST}
\end{equation}
Here, $\nu_\mathrm{eff}^{(\rm S)}$ and $v_\mathrm{m}^{(\rm S)}$ are the effective 
ionization frequency and mean velocity at SST conditions, respectively, which are 
calculated by means of the computed expansion coefficients 
$\tilde{f}_n^{(\rm S)}(v)$~\cite{Vass:2017}. 

\subsubsection{Density gradient representation}\label{Sec: BE DG}

When ionization or attachment processes become important in TOF experiments, the 
electron swarm can no longer be considered homogeneous and the electron density 
gradients become significant.

This approach to describe the electron swarm at hydrodynamic conditions (labelled 
as BE DG below) is based on an expansion of the EVDF with respect to space 
gradients of the electron density $n_\mathrm{e}$, of consecutive order. In this 
case, $f$ depends on $(\vec{r}, t)$ only via the density $n_\mathrm{e}(\vec{r}, t)$ 
and can be written as an expansion on the gradient operator $\nabla$ according to
\begin{equation}
f(\vec{r},\vec{v},t) = \sum_{j=0} F^{(j)}(\vec{v})\stackrel{j}\odot
(-\nabla)^{j}n_\mathrm{e}(\vec{r},t)\,,
\end{equation}
where the expansion coefficients $F^{(j)}(\vec{v})$ are tensors of order $j$ 
depending only on $\vec{v}$, and $\stackrel{j}\odot$ indicates a $j$-fold scalar 
product~\cite{Kumar:1980}. Note that the first coefficient $F^{(0)}(\vec{v})$ 
corresponds to the function $\hat{F}(\vec{v})$ above, for 
spatially homogeneous conditions (cf. section~\ref{Sec: BE0D}).

The expansion coefficients $F^{(j)}$ of order $j$ are obtained from a hierarchy 
of equations for each component, which all have the same structure and depend on 
the previous orders. In particular, to obtain the transport coefficients measured 
in TOF experiments, a total of five equations are required, namely for the expansion 
coefficients $F^{(0)}$, $F_z^{(1)}$, $F_T^{(1)}$, $F_{zz}^{(2)}$ and $F_{TT}^{(2)}$. 
In the present study, these equations are solved using a variant of the finite 
element method given in \cite{Segur:1983} in a $(v, \cos\theta)$ grid.

From the above expansion coefficients we obtain two sets of transport coefficients:
the \textit{flux} coefficients, neglecting the contribution of non-conservative 
processes and equivalent to those obtained by the BE 0D approach described in 
section \ref{Sec: BE0D}, and the \textit{bulk} coefficients including a contribution 
from ionization and attachment. The latter are, the \textit{bulk} drift velocity,
\begin{equation}
 W = w + \int \tilde{\nu}_\mathrm{eff}(v)F_z^{(1)}(\vec{v})\mathrm{d}\vec{v}
 \label{Eq: Wd}
\end{equation}
with $\tilde{\nu}_\mathrm{eff}(v) = vN[\sigma^{\rm i}(v)-\sigma^{\rm a}(v)]$ where 
$\sigma^{i}$ and $\sigma^{a}$ are, respectively, the ionization and attachment cross 
sections; and the longitudinal and transverse components of the diffusion tensor,
\begin{align}
 D_{\rm L} &= \int v_z F_z^{(1)}(\vec{v})\mathrm{d}\vec{v} + \int \tilde{\nu}_\mathrm{eff}(v)F_{zz}^{(2)}(\vec{v})\mathrm{d}\vec{v} \\
 D_{\rm T} &= \frac{1}{2}\left\{ \int v_T F_T^{(1)}(\vec{v})\mathrm{d}\vec{v} + \int \tilde{\nu}_\mathrm{eff}(v)F_{TT}^{(2)}(\vec{v})\mathrm{d}\vec{v} \right\} \label{Eq: DT}
\end{align}

Note that the first terms of the right-hand side of equations (\ref{Eq: Wd}-\ref{Eq: DT}) 
are the \textit{flux} component. Further details can be found in~\cite{Vass:2017}. 

The effective or apparent Townsend ionization coefficient $\alpha_\mathrm{eff}$, 
as determined in SST experiments, can be computed from the TOF parameters using 
the relation~\cite{Blevin:1984}
\begin{equation}
 \alpha_\mathrm{eff} = \frac{W}{2D_{\rm L}}-\sqrt{\left(\frac{W}{2D_{\rm L}}\right)^2
 - \frac{\nu_\mathrm{eff}}{D_{\rm L}}}, \label{eq:Blv2}
\end{equation}
which is an equivalent way of writing equation (\ref{eq:Blv1}).

\subsection{Monte Carlo technique}

In the MC simulation technique, we trace the trajectories of the electrons in the 
external electric field and under the influence of collisions. As the degree of 
ionization under the swarm conditions considered here is low, only electron-background 
gas molecule collisions are taken into account. The motion of the electrons 
with mass $m_\mathrm{e}$ between collisions is described by their equation of motion
\begin{equation}
 m_\mathrm{e}\frac{\mathrm{d}^2\vec{r}}{\mathrm{d}t^2}=-e_0\vec{E}.
 \label{Eq: Newton}
\end{equation}
The electron trajectories between collisions are determined by integrating this 
equation numerically over time steps of duration $\Delta t$ ranging between 
\SIlist{0.5;2.5}{\pico\second} for the various conditions. While this procedure 
is totally deterministic, the collisions are handled in a stochastic manner. The 
probability of the occurrence of a collision is computed after each time step, for 
each of the electrons, as
\begin{equation}
 P(\Delta t) = 1-\exp{[-N\nu\sigma^{\rm T}(v)\Delta t]}.
\end{equation}
The occurrence of a collision is determined by comparing $P(\Delta t)$ with a 
random number with a uniform distribution over the $(0,1)$ interval.
The type of collision is also selected in a random manner taking into account the 
values of the cross sections of all possible processes at the energy of the 
colliding electron. For a more detailed description see \cite{Vass:2017}.

The transport parameters (labeled as MC below) are determined as
\begin{equation}
 W = \frac{\mathrm{d}}{\mathrm{d}t}\left[\frac{\sum_{j=1}^{N_\mathrm{e}(t)}z_j(t)}
 {N_\mathrm{e}(t)}\right]\label{Eq: W_MC}
\end{equation}
and
\begin{equation}
 w = \frac{1}{N_\mathrm{e}(t)}
 \sum_{j=1}^{N_\mathrm{e}(t)}\frac{\mathrm{d}z_j(t)}{\mathrm{d}t},
\end{equation}
respectively for the \textit{bulk} and \textit{flux} drift velocities, where 
$N_\mathrm{e}(t)$ is the number of electrons in the swarm at time $t$. 
The bulk longitudinal and transverse components of the diffusion tensor are
\begin{align}
 D_{\rm L} &= \frac{1}{2}\frac{\mathrm{d}}{\mathrm{d}t}\left[\langle z^2(t)\rangle-\langle z(t)\rangle^2\right]\label{Eq: DL_MC} \\
 D_{\rm T} &= \frac{1}{4}\frac{\mathrm{d}}{\mathrm{d}t}\left[\langle x^2(t)+y^2(t)\rangle\right]\,,
\end{align}
and the effective ionization frequency is given by
\begin{equation}
 \nu_\mathrm{eff} = \frac{\mathrm{d}\ln{(N_\mathrm{e}(t))}}{\mathrm{d}t}.\label{Eq: nu_MC}
\end{equation}

Furthermore, the effective SST ionization coefficient $\alpha_\mathrm{eff}$ is 
also calculated according to relation (\ref{eq:Blv2}) using (\ref{Eq: W_MC}), 
(\ref{Eq: DL_MC}) and (\ref{Eq: nu_MC}).

All results of calculated electron swarm parameters presented in this work 
were additionally verified by independent Monte Carlo simulations and calculations 
based on multi-term solutions of the electron Boltzmann equation developed by the 
Belgrade group \cite{Dujko:2010,Dujko:2011}. For clarity, these results are not 
included in the figures shown in the next sections, but are available from the 
authors on request.

As it was already mentioned in the Introduction and is discussed in somewhat more 
detail in the next section, Monte Carlo simulations are also applied in the 
simulation of the electrons' motion in the experimental system, assisting a 
correction procedure of the experimental data.

\section{Correction of the experimental results}
\label{sec:correction}

To quantify the effect caused by the variations of the electron energy distribution
along the swarm, that in turn makes the detection sensitivity time-dependent, Monte 
Carlo simulations of 
the experimental system have been carried out  for most of the sets of conditions 
$(p, E/N)$ in the experiments. These simulations generate swarm maps, similarly 
to those measured, and a set of swarm parameters is derived from these maps via 
exactly the same fitting procedure as in the case of the experimental data. 
The transport parameters and ionization frequencies obtained from the simulations 
of the setup are compared with those obtained from kinetic swarm calculations based 
on the solution of the electron Boltzmann equation, where the same electron collision 
cross section sets are used. Good agreement between the two sets of swarm parameters 
implies that the assumption made in the fitting of the experimental data, i.e.\ the use 
of the theoretical form (\ref{eq:n}) of $n_{\rm e}(z,t)$ as a fit to the measured 
data, is acceptable. In contrast, strong deviations indicate that this assumption 
is not applicable for the given condition. We note that no experimental data are 
involved in this procedure.

\begin{figure}[h]
 \centering
 \includegraphics[width=0.9\linewidth]{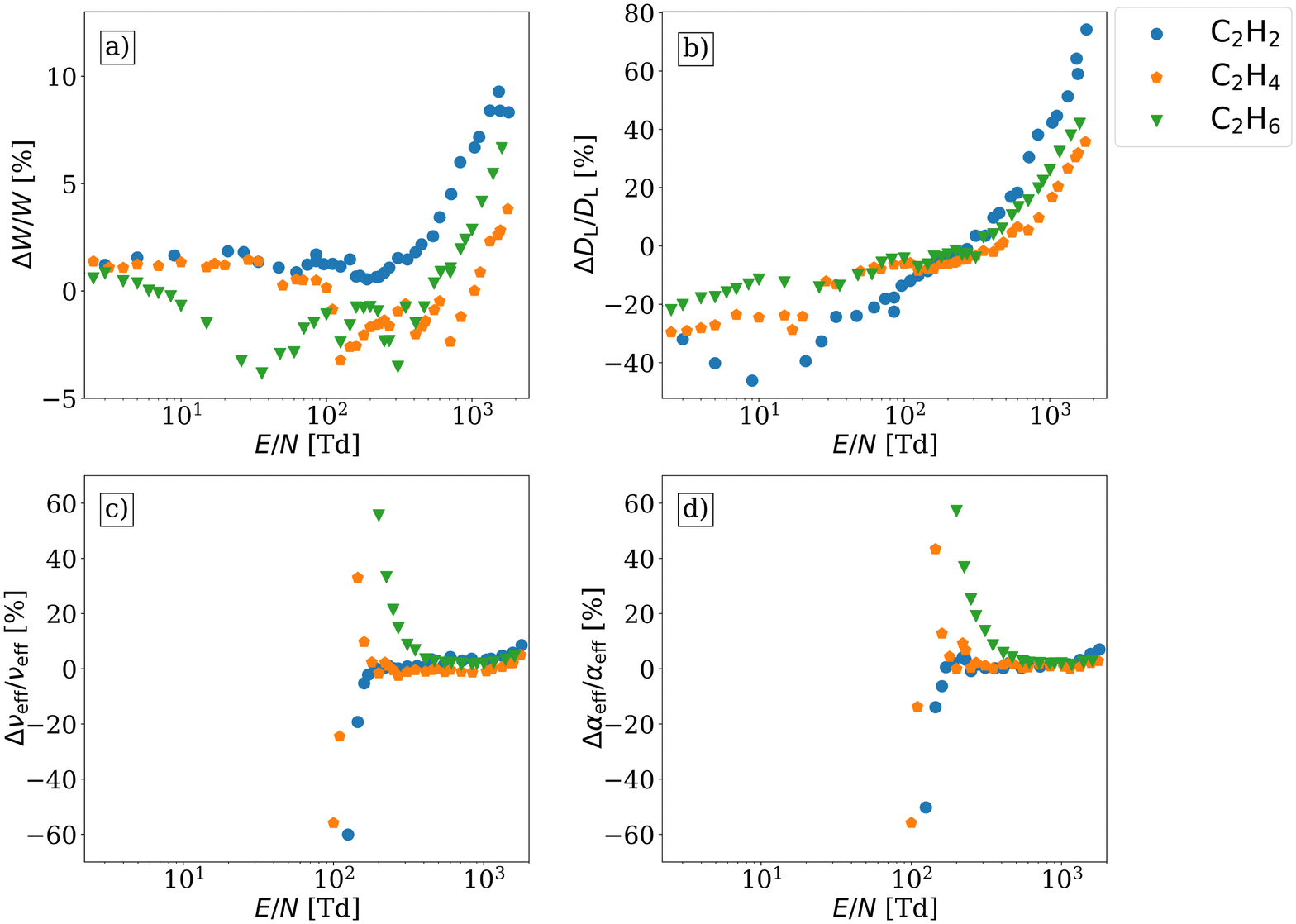}
 \caption{Deviations of the results between the swarm parameters obtained 
 from the simulations of the experimental systems vs. the theoretical values. 
 (a) Bulk drift velocity, (b) longitudinal component of the diffusion tensor, (c) effective 
 ionization frequency, and (d) effective SST ionization coefficient. Applying 
 these correction factors to the experimental results leads to the set of 
 'Corrected' transport coefficients.}
 \label{fig:validation}
\end{figure}

Results of this procedure for each of the gases and for the whole domain of $E/N$ 
are presented in figure~\ref{fig:validation}. The panels correspond to the swarm 
parameters $W$, $D_{\rm L}$, $\nu_{\rm eff}$, and $\alpha_{\rm eff}$, respectively, 
and show the differences of each parameter derived by the simulation of the 
experimental system with respect to its theoretical value obtained from the BE solution.

In the case of the bulk drift velocity (figure \ref{fig:validation}(a), the error 
is in the few \% range for most of the conditions, and it approaches 
$\approx$ \SI{10}{\percent} at the highest $E/N$ values. This indicates that the 
determination of the bulk drift velocity values from the experimental data is quite
relyable.

The situation turns out to be much worse for the longitudinal component of the diffusion 
tensor (figure~\ref{fig:validation}(b)). Here, the error ranges from 
$\approx-$\SI{40}{\percent} to $\approx+$\SI{80}{\percent}, depending on $E/N$. 
The $D_{\rm L}$ data can be considered to be acceptably accurate at intermediate 
$E/N$ values only. The much larger error of $D_\mathrm{L}$ with respect to that 
of $W$ can be explained by the fact that the distribution of the average electron 
energy along the swarm is 
inhomogeneous. In the close vicinity of the maximum of the spatial distribution 
of the electron density, the variation of the average energy along the swarm is 
comparatively small. However, by moving away from this maximum, the spatial 
variations of the average energy along the swarm increase. As the drift velocity 
is primarily determined by the position of the maximum of the spatial profile of 
the electron density while the diffusion is predominantly determined 
by the width of this distribution, it is clear that the width of the distribution 
is more affected by non-uniform sensitivity of the detector with respect to the 
average electron energy than the position of the maximum.

Regarding the effective ionization frequency (figure~\ref{fig:validation}(c)) and 
the strongly related SST ionization coefficient (figure~\ref{fig:validation}(d)), 
we observe small errors at high $E/N$ values, where ionization is appreciable. The 
error, on the other hand, grows high when $E/N$ approaches $\approx$ \SI{100}{Td}, 
where both $\nu_{\rm eff}$ and $\alpha_{\rm eff}$ drop rapidly.

\section{Results and discussion}
\label{Sec:results}

The electron swarm parameters have been measured in a wide range of the 
reduced electric field, between \SIlist{1;1790}{Td} at a gas temperature $T$ of 
\SI{293}{K}. In the following, results of our measurements are presented for 
the three hydrocarbon gases \ce{C2H2}, \ce{C2H4}, and \ce{C2H6}. Besides the 
transport parameters and ionization coefficients resulting from the experiments via 
the fitting procedure described in section \ref{sec:intro}, we also present the 
corrected values of these data resulting from the procedure introduced in 
section~\ref{sec:correction}. For each swarm 
parameter, we compare the present measured data with previous 
experimental results and with the results of the kinetic computations based on the 
solution of the electron Boltzmann equation or on MC simulations, obtained with the 
selected electron collision cross sections. The results for the \textit{flux} parameters 
obtained by methods \mbox{BE 0D}, \mbox{BE DG} and MC overlap, and so do the \textit{bulk} 
parameters obtained from the \mbox{BE DG} and MC methods.

\subsection{Electron mobility}

\begin{figure}[tp]
 \centering
 \includegraphics[width=0.60\linewidth]{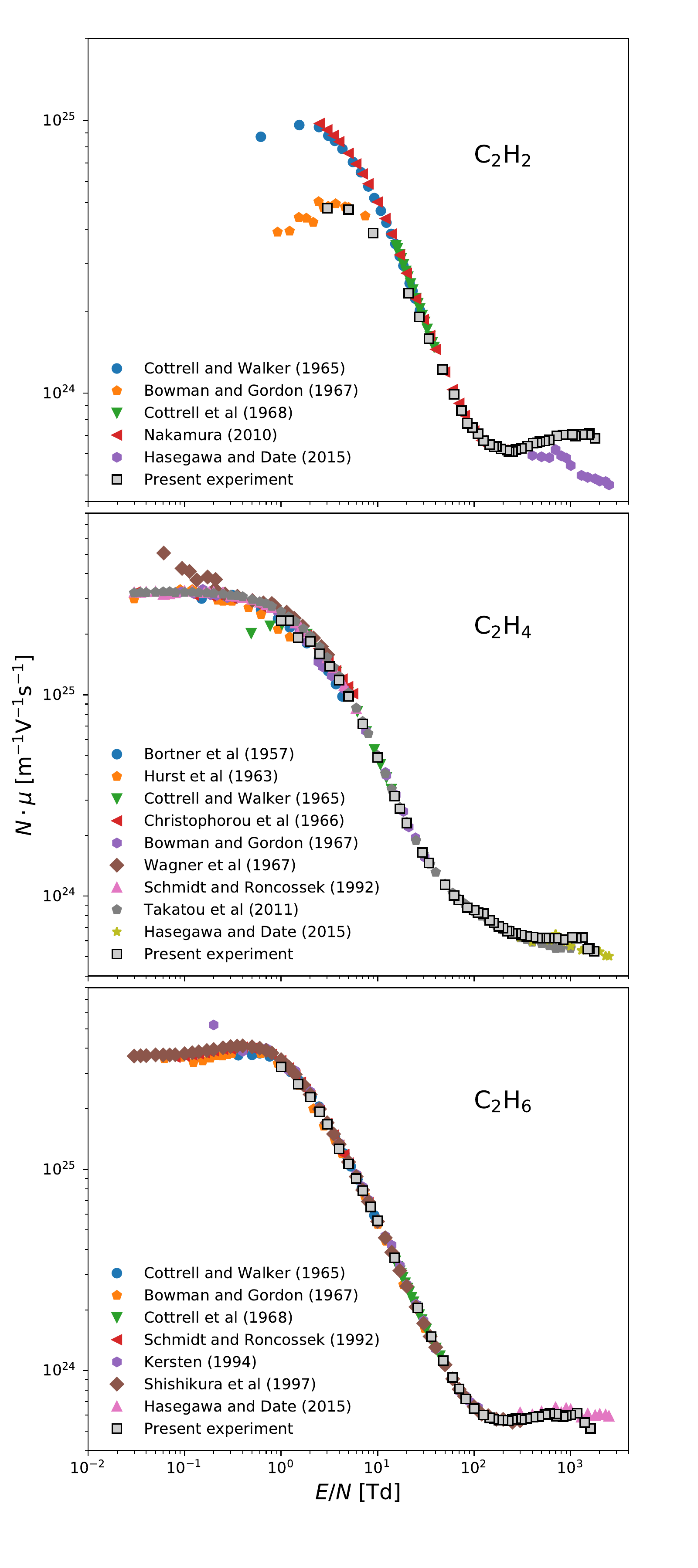}
 \caption{Mobility in \ce{C2H2}, \ce{C2H4} and \ce{C2H6} obtained from drift velocity 
 results: Bortner \textit{et al}~\cite{Bortner:1957}, Hurst \textit{et al}~\cite{Hurst:1963}, 
 Cottrell and Walker~\cite{Cottrell:1965}, Christophorou \textit{et al}~\cite{Christophorou:1966}, Bowman and Gordon~\cite{Bowman:1967}, Wagner \textit{et al}~\cite{Wagner:1967}, 
 Cottrell \textit{et al}~\cite{Cottrell:1968},  Schmidt and Roncossek~\cite{Schmidt:1992}, 
 Kersten~\cite{Kersten:1994},  Shishikura \textit{et al}~\cite{Shishikura:1997}, 
 Nakamura~\cite{Nakamura:2010}, Takatou \textit{et al}~\cite{Takatou:2011},
 Hasegawa and Date~\cite{Hasegawa_Date:2015} and present measurements. The figures 
 share the same \textit{E/N} scale. 'Present experiment' corresponds to the 
 uncorrected experimental data. The corrected data are not shown here because of 
 the small correction factors for the bulk drift velocity and the mobility.}
 \label{fig: C2Hm-Mob_Exp}
\end{figure}

We start by comparing the values of the reduced mobility, $N\,\mu$, derived from 
the bulk drift velocity, with previous experimental data for the three hydrocarbon 
gases in figure~\ref{fig: C2Hm-Mob_Exp}. Our experimental results for the (uncorrected
and corrected) bulk drift velocity are compiled in table \ref{tab:WC2Hnwithneq246} 
in \ref{App A}. We estimate the maximum experimental error of these values to be 
around \SI{6}{\percent}.

Except for the high values of $E/N$, our measured \textit{bulk} drift velocity and 
mobility results are in excellent agreement with all previous results.
In \ce{C2H2}, however, at low $E/N$ we find two distinct sets of results: the present 
results are consistent with the measurements of Bowman and Gordon~\cite{Bowman:1967}, 
while the results of Cottrell and Walker~\cite{Cottrell:1965} are in accordance 
with those of Nakamura~\cite{Nakamura:2010}. Note that the latter results were used 
to obtain the recommended electron collision cross sections for \ce{C2H2}~\cite{Song:2017} 
used in the present modelling and simulation. At high $E/N$ the present results 
deviate from those of Hasegawa and Date~\cite{Hasegawa_Date:2015} in \ce{C2H2} 
and \ce{C2H4}. 
However the latter results are obtained from the mean arrival-time velocity defined 
in~\cite{Kondo:1990} and are not easily comparable with the present TOF results in 
the presence of reaction processes.

\begin{figure}[tp]
 \centering
 \includegraphics[width=0.9\linewidth]{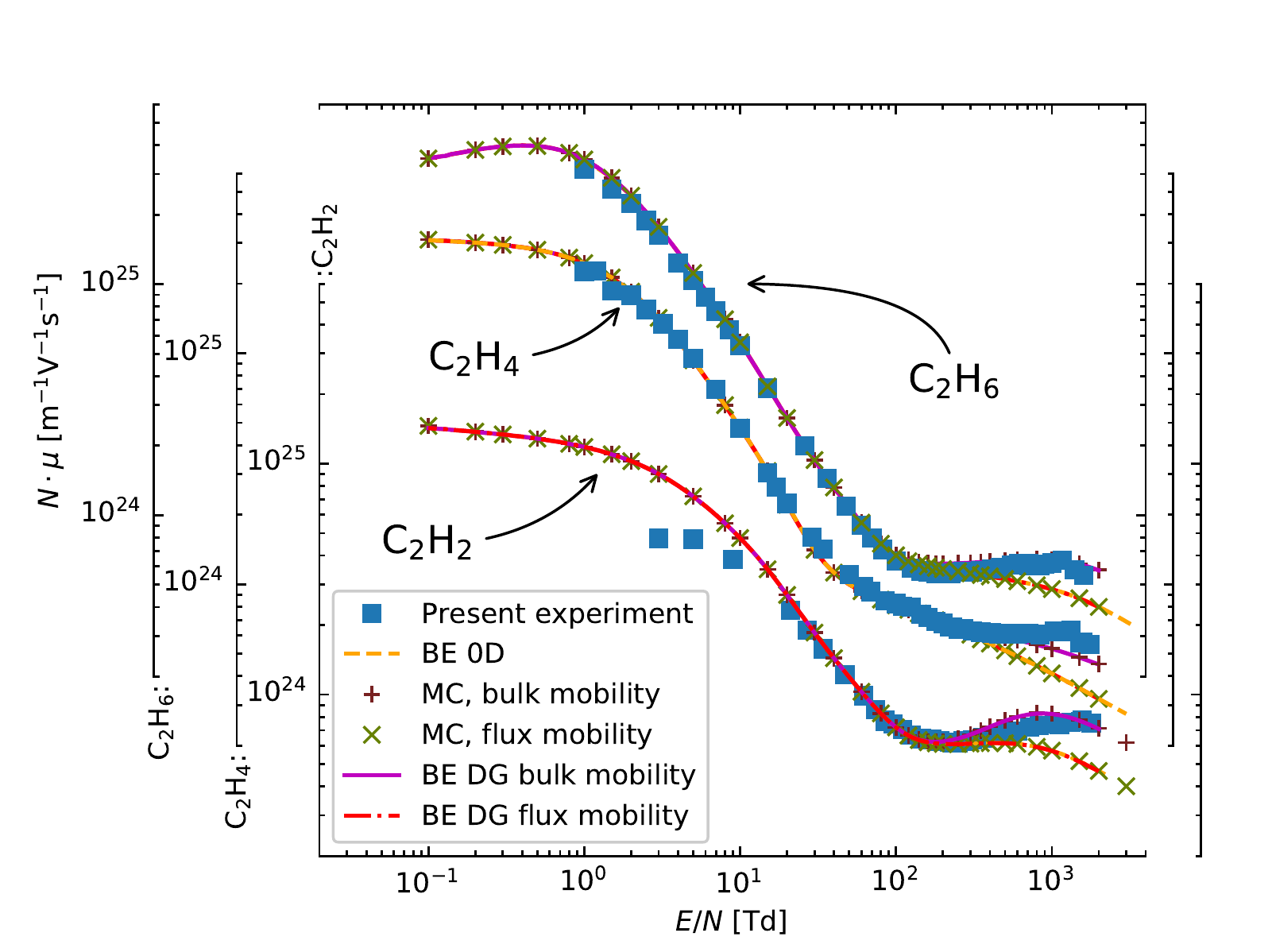}
 \caption{Mobility in \ce{C2H2}, \ce{C2H4} and \ce{C2H6}: present experiment and 
 modelling results. The results and $N\mu$ scale for \ce{C2H4} and \ce{C2H6} 
 have been shifted. 'Present experiment' corresponds to the uncorrected
 experimental data. The corrected data are not shown here because of the small 
 correction factors for the bulk drift velocity / mobility.}
 \label{fig: C2Hm-Mob_Mod}
\end{figure}

In figure \ref{fig: C2Hm-Mob_Mod} we compare the results of the present measurements 
with the kinetic computation results. In this figure the \textit{E/N} scale is common 
to the three gases but the $N\mu$ scale and data for \ce{C2H4} and \ce{C2H6} have been 
shifted upwards to avoid overlapping of the curves.
Above \SI{200}{Td} the contribution of non-conservative processes becomes visible 
and the mobility results are split into a \textit{bulk} branch (for MC and \mbox{BE DG} 
\textit{bulk} mobilities and the present measurements) and \textit{flux} values 
(respectively for \mbox{BE 0D}, MC and \mbox{BE DG} \textit{flux} mobilities).
Here our measured data show some differences to the MC and \mbox{BE DG} bulk results 
for all three gases. In case of \ce{C2H2}, as the electron collision cross sections 
used are based on the swarm results of Nakamura~\cite{Nakamura:2010}, the modelling 
results deviate from the present experimental results below \SI{10}{Td}. Note that 
below \SI{3}{Td} the modelling results also deviate from the measurements of Bowman 
and Gordon~\cite{Bowman:1967} as well as of Cottrell and Walker~\cite{Cottrell:1965} 
in figure~\ref{fig: C2Hm-Mob_Exp}. 

\subsection{Diffusion tensor}

\begin{figure}[tp]
 \centering
 \includegraphics[width=0.62\linewidth]{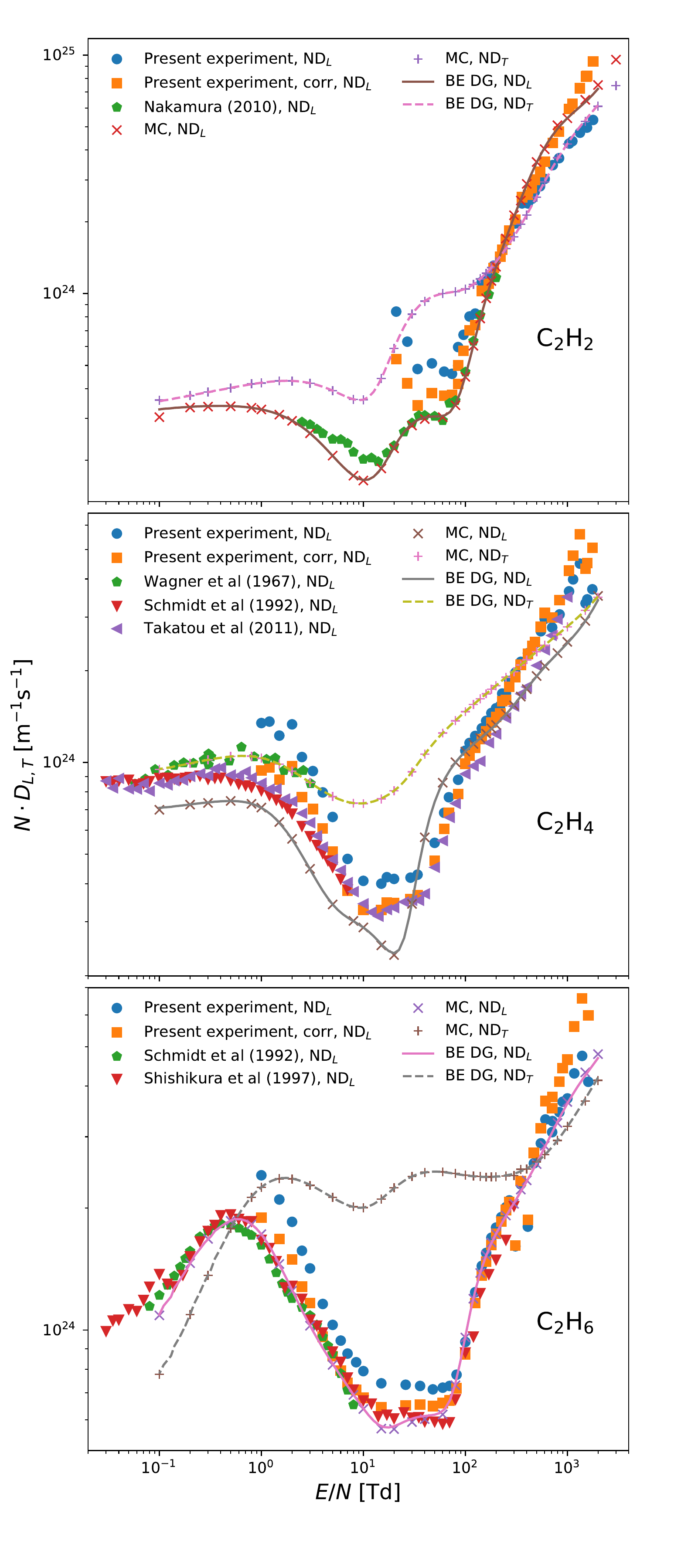}
 \caption{Longitudinal and transverse \textit{bulk} components of the diffusion 
 tensor in \ce{C2H2}, \ce{C2H4} and \ce{C2H6}. Experimental results: present 
 experiment, Wagner \textit{et al}~\cite{Wagner:1967}, Schmidt and Roncossek~\cite{Schmidt:1992},
 Shishikura \textit{et al}~\cite{Shishikura:1997}, Nakamura~\cite{Nakamura:2010}, 
 Takatou \textit{et al}~\cite{Takatou:2011}. Modelling results: MC and BE DG 
 ($ND_{\rm L}$ and $ND_{\rm T}$).  The figures share the same $E/N$ scale. The 
 panels show both the uncorrected  and corrected experimental results of this study.}
 \label{fig: C2Hm-ND}
\end{figure}

The present experimental results for the gas number density times the longitudinal 
component of the diffusion tensor, $N D_{\rm L}$, for \ce{C2H2}, \ce{C2H4} 
and \ce{C2H6} are given in \ref{App A}, table \ref{tab:NDLC2Hnwithneq246}.
They are shown in figure \ref{fig: C2Hm-ND} together with previously measured data 
as well as with the kinetic computation values for the \textit{bulk} longitudinal 
and transverse components of the diffusion tensor for each gas. 
The present measured values of $N D_{\rm L}$ exhibit larger scattering, which is
explained by the higher uncertainty of the determination of $D_{\rm L}$ in the 
experiments ($\approx\SI{10}{\percent}$) compared to that of the drift velocity. 

Above \SI{100}{Td} there is reasonable agreement of the present measurements with 
previous experimental data and the modelling results, for the three gases. Below 
\SI{100}{Td} however, the present measurements evidence the same qualitative 
behaviour but are systematically above previous measurements. Note that the 
application of the correction procedure, detailed in section~\ref{sec:correction}, 
to our experimental results leads to much better agreement with previously measured 
data, in particular for \ce{C2H4} and \ce{C2H6}.

The modelling results for $D_{\rm L}$ in \ce{C2H2} and \ce{C2H4} below \SI{2}{Td} 
and \SI{5}{Td}, respectively, also deviate from all experimental results indicating 
that the corresponding cross section sets require improvement. 
In each of the three gases, the values of the transverse component of the diffusion 
tensor, $D_{\rm T}$, obtained by the kinetic computations, are very different 
from the longitudinal component, $D_{\rm L}$. The measurement of data of this component 
can provide additional tests for the fitting of the electron collision cross sections.

\subsection{Effective ionization frequency and SST ionization coefficient}

\begin{figure}[tp]
 \centering
 \includegraphics[width=0.9\linewidth]{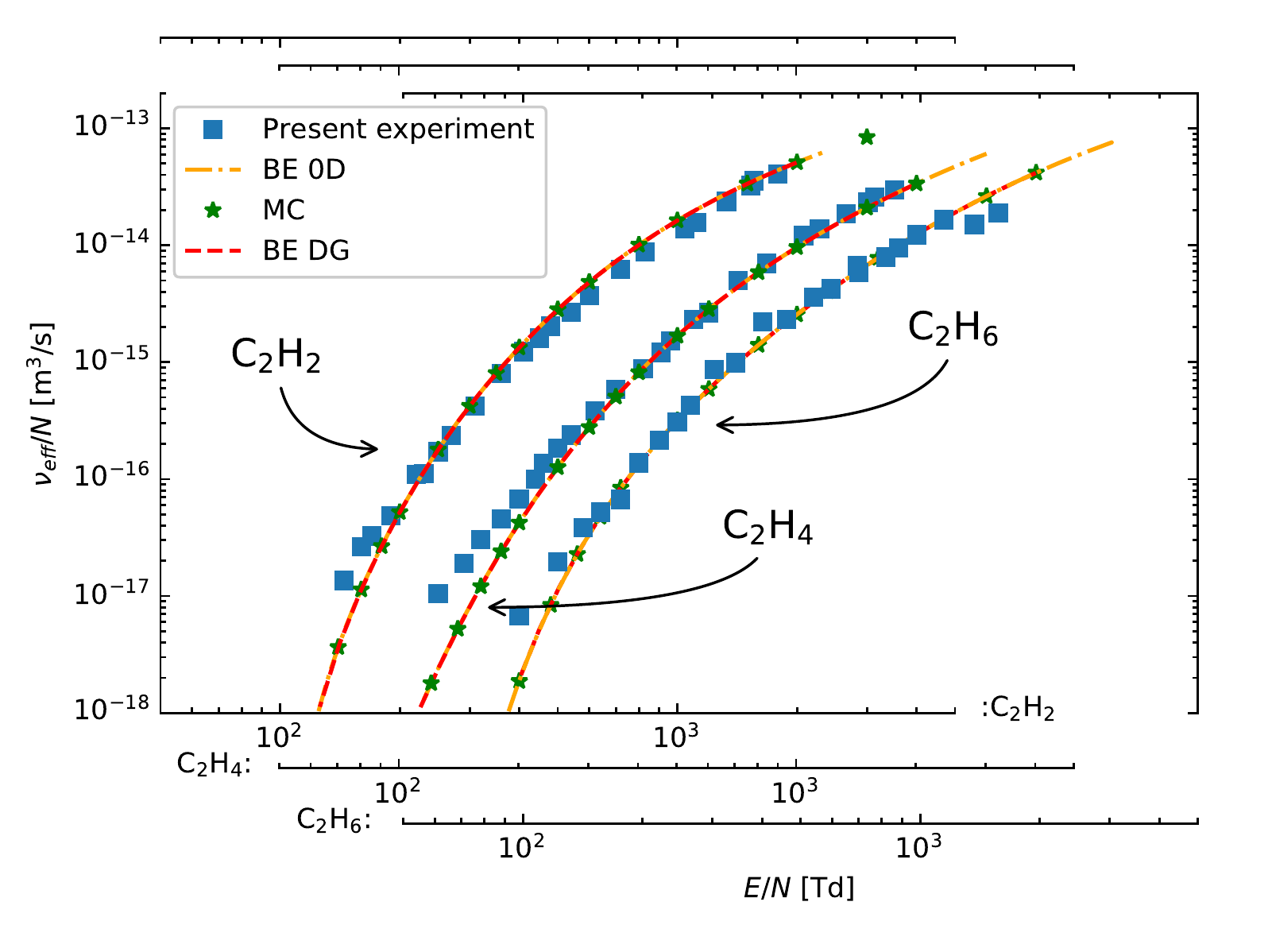}
 \caption{Reduced effective ionization frequency in \ce{C2H2}, \ce{C2H4} and \ce{C2H6}: 
 present experiment and modelling results. The results and $E/N$ scale for \ce{C2H4} 
 and \ce{C2H6} are shifted horizontally. 'Present experiment' corresponds to the 
 uncorrected data.}
 \label{fig: C2Hm-Nui_eff}
\end{figure}

The experimental and modelling results for the reduced effective ionization 
frequency, $\nu_\mathrm{eff}/N$, for the three gases studied are displayed 
in figure \ref{fig: C2Hm-Nui_eff}.
To our best knowledge this is the first report of $\nu_\mathrm{eff}$ in these three 
gases for an extended range of $\SI{100}{Td} \le E/N \le \SI{1790}{Td}$, for which 
the estimated experimental error of the data is $\le \SI{8}{\percent}$.
Our measured 
data are also listed in \ref{App A}, table~\ref{tab:nueffpNC2Hnwithneq246}. 
In order to accommodate the results on the same figure, all gases share a same 
$\nu_\mathrm{eff}/N$ axis but the $E/N$ scales for \ce{C2H4} and \ce{C2H2} have 
been shifted to the right.

Good agreement between our measured and calculated results is generally found for 
$E/N$ values larger than about \SI{200}{Td}, indicating that the electron
collision cross section sets for the three gases are reasonably well adapted to 
allow for an appropriate determination of the rate coefficients for ground state
ionization. 
Certain differences are obvious for lower $E/N$ values. These differences seem 
to result from the measurement and/or, more likely, from the fitting procudure 
(see figure~\ref{fig:validation}).

\begin{figure}[tp]
 \centering
 \includegraphics[width=0.65\linewidth]{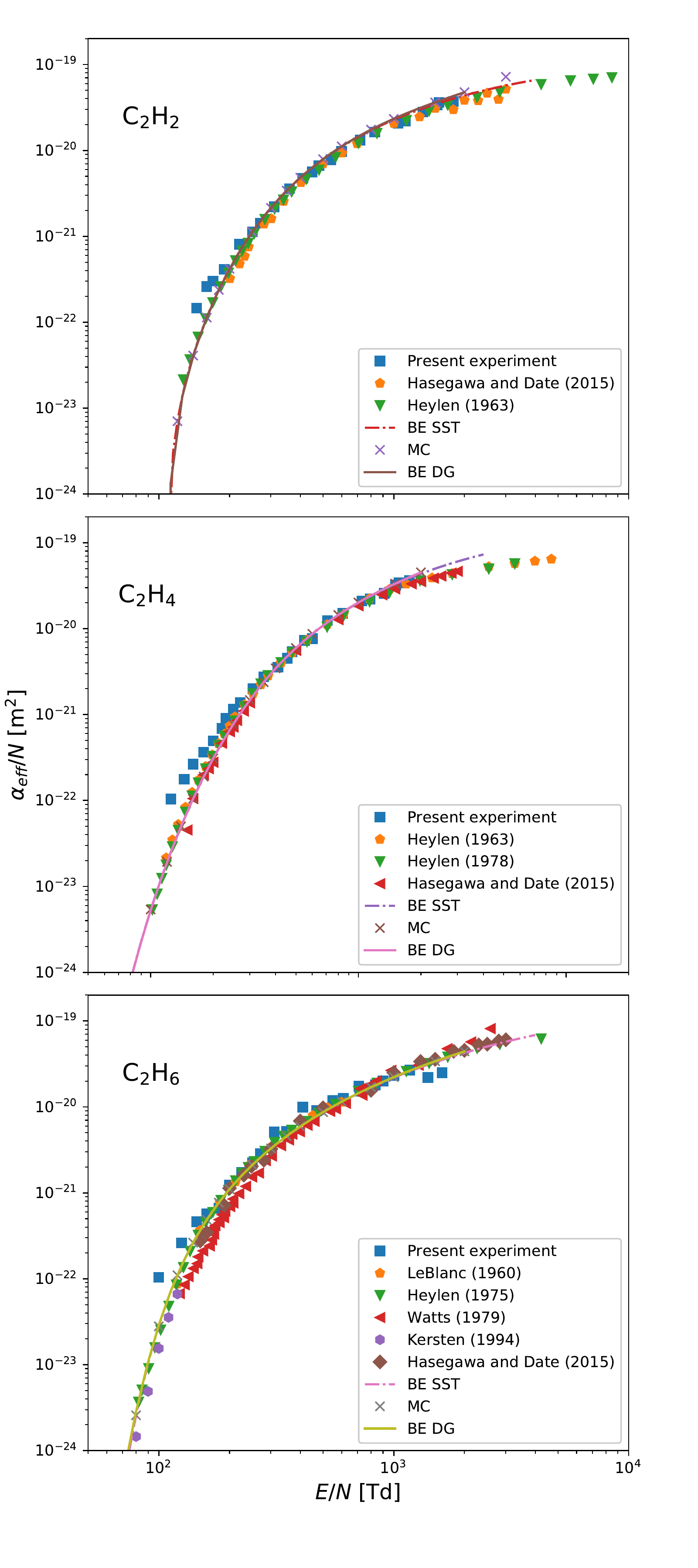}
 \caption{Reduced effective Townsend ionization coefficient in \ce{C2H2}, \ce{C2H4} 
 and \ce{C2H6}. Experimental results: present experiment, 
 Heylen~\cite{Heylen:1963,Heylen:1975},  
 Watts and Heylen~\cite{Watts:1979}, Kersten~\cite{Kersten:1994} and Hasegawa and 
 Date~\cite{Hasegawa_Date:2015}. Modelling results: BE SST, MC and BE DG. 
 'Present experiment' corresponds to the uncorrected data.}
 \label{fig: C2Hm-alpha}
\end{figure}

Our experimental data for the reduced effective SST ionization coefficient, 
$\alpha_\mathrm{eff}/N$, obtained using equation~(\ref{eq:Blv1}), are compared 
with previous measurements and the kinetic computation results in figure 
\ref{fig: C2Hm-alpha}. As $\alpha_\mathrm{eff}$ is derived from the set of 
parameters $\{W, D_{\rm L}, \nu_\mathrm{eff}\}$, these results have a higher 
uncertainty than $\nu_\mathrm{eff}$ with an estimated experimental error of 
$\le$ \SI{10}{\percent}. Notice that the kinetic computation results using method 
\mbox{BE SST} do not include the approximations involved in equation (\ref{eq:Blv1}), 
but are directly obtained by solving the electron Boltzmann equation at SST 
conditions according to (\ref{eq:alphaeffBESST}). 
In this respect, their comparison with the \mbox{BE DG} and MC results can indicate 
the range of validity of equation~(\ref{eq:Blv1}) or (\ref{eq:Blv2}).  
Our experimental results of $\alpha_\mathrm{eff}/N$ are compiled in \ref{App A}, 
table~\ref{tab:alphaeffpNC2Hnwithneq246} as well.

Except for the low values of $E/N$, our results for the effective SST 
ionization coefficient are in excellent agreement with all previous results and 
the kinetic computations. At values close to the threshold, however, the present 
results are higher than previous measurements. 
Notice that Kersten's effective Townsend ionization coefficient was measured under 
TOF conditions and corresponds to $\nu_\mathrm{eff}/W$~\cite{Kersten:1994}. 
Thus, it represents the effective SST ionization coefficient $\alpha_\mathrm{eff}$ 
according to~(\ref{eq:Blv1}) only in the absence of diffusion, i.e., $D_{\rm L}=0$.

\subsection{Effect of the vibrationally excited population}\label{Sec: VibEx}

The cross sections sets used above were obtained considering only electron collisions 
with the ground state of the molecules. However, as polyatomic molecules have multiple 
vibrational modes and these modes can be degenerated, in these gases we can find 
a significant fraction of molecules in thermally excited 
vibrational states at room temperature. In addition to their contribution to energy 
losses due to elastic, exciting, ionizing and attaching collision processes, these 
excited states contribute to electron energy gains due to superelastic 
collisions and influence the EVDF and transport parameters, mainly at low to 
medium $E/N$ field values. The importance of their effect increases with the 
energy associated with the collision and the fractional population of thermally 
excited states with that energy. This population, however, decreases exponentially 
with energy. From the combination of these two factors, the effect on the EVDF 
should be maximum for a given energy value.

Taking into account the equations for the fractional populations and statistical 
weights of polyatomic molecules in 
\ref{App B}, we can estimate the populations 
of the different states of these gases.

\begin{description}
 \item[Acetylene] has five vibrational modes, with the two bending modes ($v_4$ 
 and $v_5$) double degenerated and with energies of, respectively, \SI{0.075}{eV} 
 and \SI{0.0905}{eV} \cite{Sh1972NBS}. At a gas temperature of \SI{293.15}{K}, 
 the vibrational states with fractional population above \SI{0.1}{\percent} are 
 indicated in table~\ref{tab:FracPop}. 
 At this temperature only around \SI{85}{\percent} of the acetylene molecules are 
 in the ground state and the vibrational population in excited states of modes 
 $v_4$ and $v_5$ is significant.

\begin{table*}[htp]
 \caption{Fractional population of the first vibrational levels of \ce{C2H2} 
 at \SI{293.15}{K}.\\}
 \label{tab:FracPop}
 \centering
\begin{tabular}{cccSS} \toprule
Vibr. state & short notation & g & \textrm{Energy (eV)} & \textrm{Frac. pop. $(\%)$}\\ \midrule
(00000) & $v_0$ & 1 & 0.0 & 85.37\\
(10000) & $v_1$ & 1 & 0.421 & 5.5e-6\\
(01000) & $v_2$ & 1 & 0.245 & 5.3e-3\\
(00100) & $v_3$ & 1 & 0.411 & 8.3e-6\\
(00010) & $v_4$ & 2 & 0.075 & 8.47\\
(00020) &       & 3 & 0.150 & 0.63\\
(00001) & $v_5$ & 2 & 0.0905 & 4.75\\
(00002) &       & 3 & 0.180 & 0.20\\
(00011) & $v_4+v_5$ & 4 & 0.165 & 0.47 \\ \bottomrule
\end{tabular}
\end{table*}

 \item[Ethylene:] In contrast to \ce{C2H2}, none of the twelve ethylene vibrational 
 modes \cite{Sh1972NBS} is degenerated, where the lowest threshold energy for 
 vibrational excitation to $v_{10}$ is \SI{0.102}{eV} and, at the same temperature, 
 more than \SI{95}{\percent} of the molecules are in the ground state. 
 
 \item[Ethane:] All the degenerated vibrational modes of ethane \cite{Sh1972NBS} 
 have energies above \SI{0.15}{eV} and at room temperature their fractional 
 population is small. Overall, however, only \SI{73}{\percent} of ethane molecules 
 are in the ground state as mode $v_4$ has an excitation energy of only \SI{0.036}{eV}. 
 Molecules in the two first excited vibrational states of this mode represent 
 \SI{22}{\percent} of the total. On the other hand, as the excitation energy of 
 the $v_4$ mode transitions is very small, the effect on the EVDF and transport 
 parameters is also small. 

\end{description}

Of the three gases analysed, the impact of the thermally excited vibrational 
population on the EVDF should be largest in \ce{C2H2}. The vibrational excitation 
cross section set for \ce{C2H2}~\cite{Song:2017} is also more complete than the 
vibrational cross section sets for \ce{C2H4} and \ce{C2H6} used in this study. For 
these reasons we study the effect of the thermally excited vibrational states only 
for acetylene.

Our goal is to single out the contribution of the vibrationally excited molecules 
due to superelastic collisions and we will change the electron collision cross 
sections in such a 
way that, if we neglect these collisions, we obtain the same results as before.
Starting from the recommended cross section set for ethylene~\cite{Song:2017}, we 
introduce the following modifications: \\
a) We split the lumped cross sections for the vibrational excitation of modes 
$v_1/v_3$ and $v_4/v_5$ into individual cross sections for each modes, with a value 
of half of the original cross section. That is 
 $\sigma_{v_1} = \sigma_{v_3} =\frac{1}{2}\sigma_{v_1/v_3}$ and 
 $\sigma_{v_4} = \sigma_{v_5} =\frac{1}{2}\sigma_{v_4/v_5}$. \\
b) The threshold for the excitation of modes $v_1$ and $v_3$ and of modes $v_4$ and 
$v_5$ is set at the same value as before of, respectively, \SI{0.411}{eV} and 
\SI{0.0905}{eV}. \\
c) We assume that all molecules are in one of the three states $(00000)$, $(00010)$ 
and $(00001)$, with the fractional population, $\delta$, of the last two states 
in thermal equilibrium with the gas and the ground state fraction given by 
$\delta_{00000}=(1-\delta_{00010}-\delta_{00001})$. \\
d) We consider the following vibrational excitation processes for electron collisions 
with the ground state $(00000)$:
\[\begin{array}{lcl} 
\ce{e} + \ce{C2H2}(00000) & \to & \ce{e} + \ce{C2H2}(10000)\\
\ce{e} + \ce{C2H2}(00000) & \to & \ce{e} + \ce{C2H2}(01000)\\
\ce{e} + \ce{C2H2}(00000) & \to & \ce{e} + \ce{C2H2}(00100)\\
\ce{e} + \ce{C2H2}(00000) & \leftrightarrow & \ce{e} + \ce{C2H2}(00010)\\
\ce{e} + \ce{C2H2}(00000) & \leftrightarrow & \ce{e} + \ce{C2H2}(00001)\\
\ce{e} + \ce{C2H2}(00000) & \leftrightarrow & \ce{e} + \ce{C2H2}(00011)
\end{array}\]
where reactions with double-arrows include superelastic collisions.\\
f) We additionaly include the following vibrational excitation processes on 
collisions with states $(00010)$ and $(00001)$:
\[\begin{array}{lcl} 
\ce{e} + \ce{C2H2}(00010) & \to & \ce{e} + \ce{C2H2}(10010)\\
\ce{e} + \ce{C2H2}(00010) & \to & \ce{e} + \ce{C2H2}(01010)\\
\ce{e} + \ce{C2H2}(00010) & \to & \ce{e} + \ce{C2H2}(00110)\\
\ce{e} + \ce{C2H2}(00010) & \leftrightarrow & \ce{e} + \ce{C2H2}(00020)\\
\ce{e} + \ce{C2H2}(00010) & \leftrightarrow & \ce{e} + \ce{C2H2}(00011)\\
\ce{e} + \ce{C2H2}(00010) & \to & \ce{e} + \ce{C2H2}(00021)
\end{array}\]
and
\[\begin{array}{lcl} 
\ce{e} + \ce{C2H2}(00001) & \to & \ce{e} + \ce{C2H2}(10001)\\
\ce{e} + \ce{C2H2}(00001) & \to & \ce{e} + \ce{C2H2}(01001)\\
\ce{e} + \ce{C2H2}(00001) & \to & \ce{e} + \ce{C2H2}(00101)\\
\ce{e} + \ce{C2H2}(00001) & \leftrightarrow & \ce{e} + \ce{C2H2}(00011)\\
\ce{e} + \ce{C2H2}(00001) & \leftrightarrow & \ce{e} + \ce{C2H2}(00002)\\
\ce{e} + \ce{C2H2}(00001) & \to & \ce{e} + \ce{C2H2}(00012)
\end{array}\]
adopting for these processes the same cross sections as the corresponding 
excitations from the ground state.\\
e) We further assume that the electron collision cross sections for momentum 
transfer, electronic excitation, ionization and attachment with the vibrational 
states $(00010)$ and $(00001)$ are the same as for state $(00000)$.

Note that if we neglect superelastic collisions, the EVDF and swarm parameters 
obtained with these modified cross sections and electron collision reactions are 
exactly the same as with the original set~\cite{Song:2017} and are independent 
of the fractional population of levels $(00010)$ and $(00001)$.

\begin{figure}[tp]
 \centering
 \includegraphics[width=0.65\linewidth]{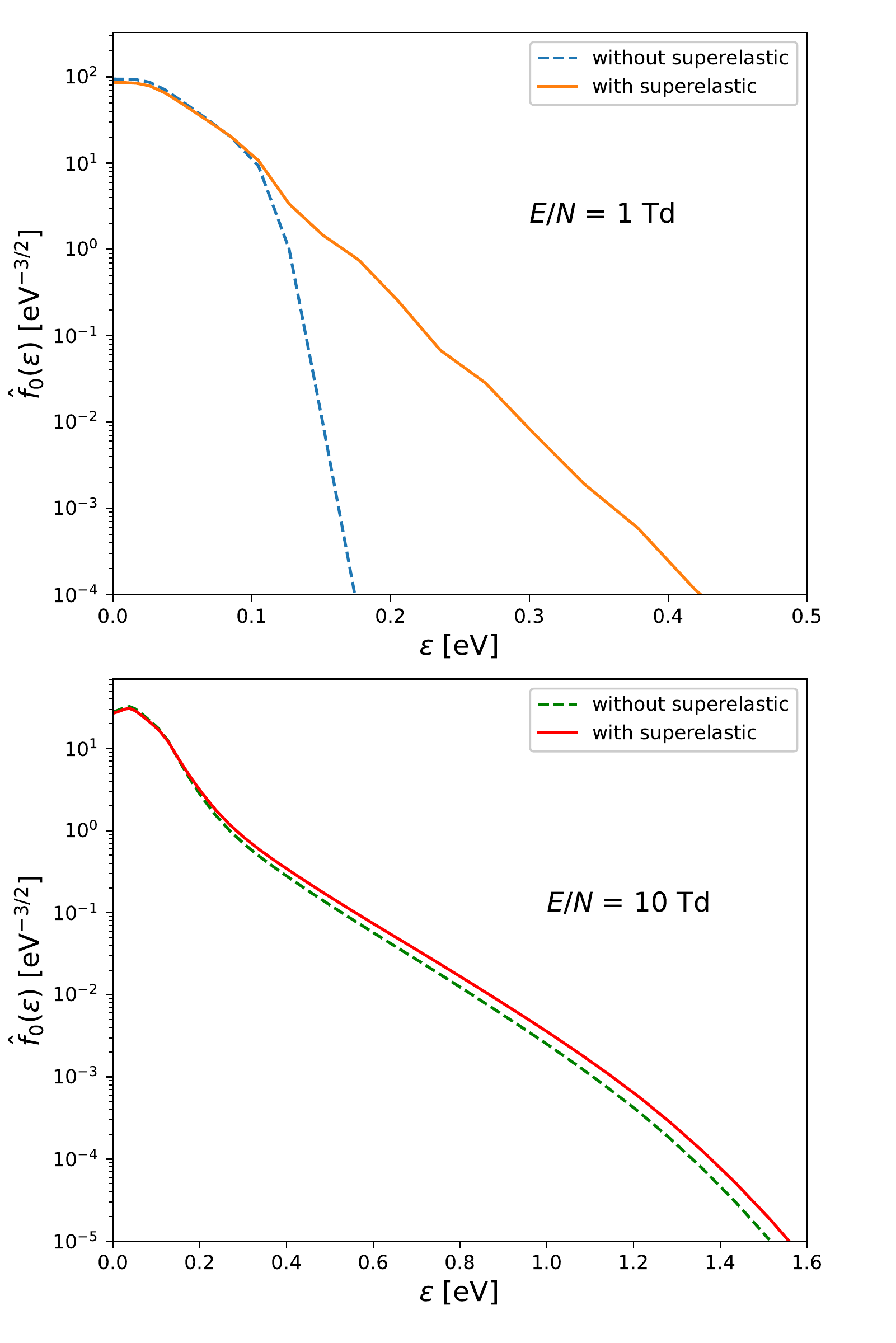}
 \caption{Isotropic component of the EVDF in \ce{C2H2} at \SI{293.15}{K} for 
 \SIlist{1;10}{Td}, with and without superelasic collision processes included.}
 \label{fig: C2H2-eedf}
\end{figure}

The influence of superelastic collisions is ilustrated in figure \ref{fig: C2H2-eedf} 
which shows the isotropic component $\hat{f}_0(\epsilon)$ of the EVDF as a function 
of the electron kinetic energy, $\epsilon=m_\mathrm{e}v^2/2$, calculated at $E/N$ 
values of \SIlist{1;10}{Td}, respectively, with and without the inclusion of 
superelastic processes. 
Pronounced differences between the corresponding isotropic distributions 
$\hat{f}_0(\epsilon)$ are found at $E/N = \SI{1}{Td}$, while the impact of 
superelastic electron collision processes is comparatively small  at \SI{10}{Td}. 
This finding is not only reflected by the isotropic distribution but also by different 
macroscopic properties.

\begin{figure}[tp]
 \centering
 \includegraphics[width=0.65\linewidth]{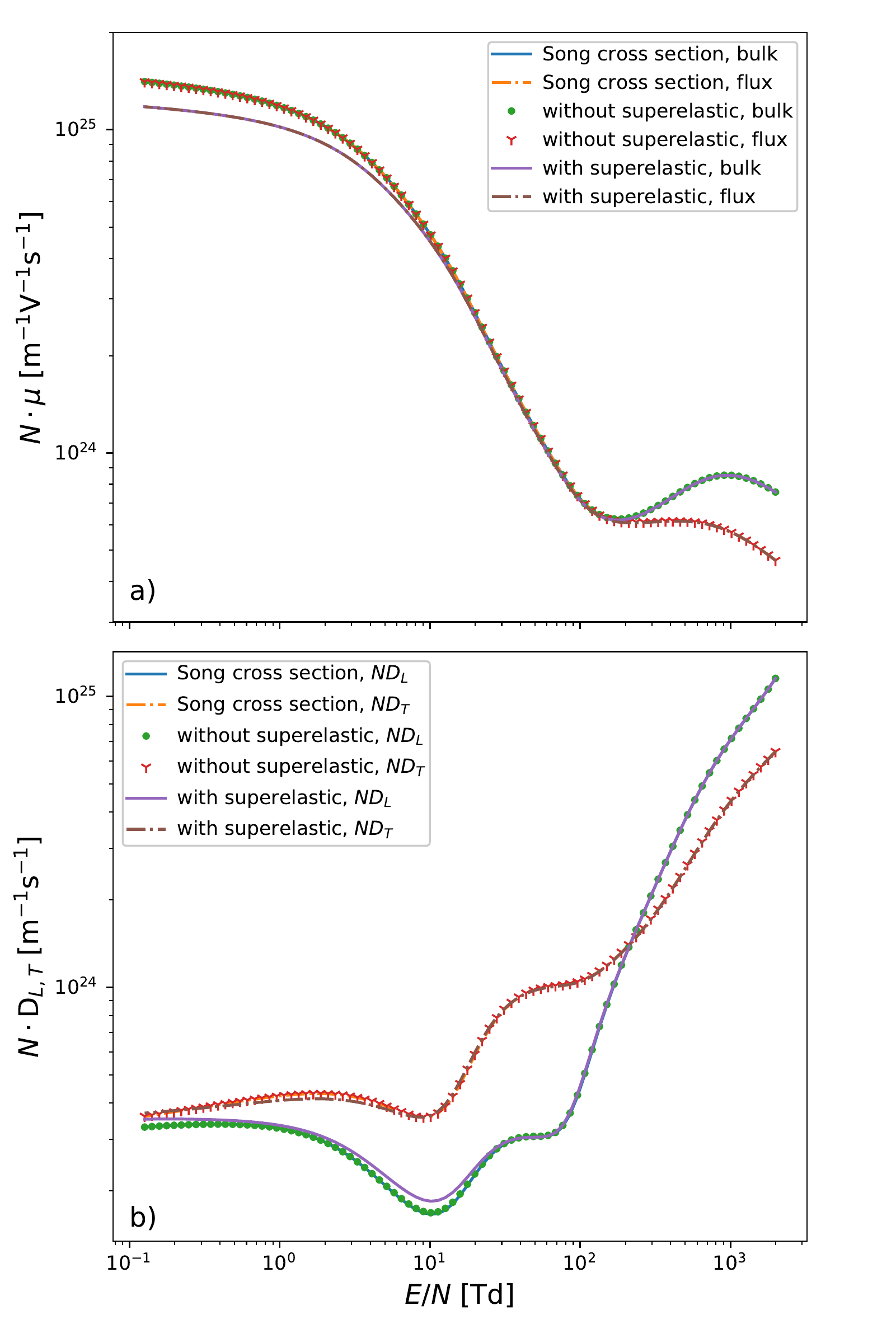}
 \caption{(a) Mobility and longitudinal and (b) transverse \textit{bulk} components of the 
 diffusion tensor in \ce{C2H2} at \SI{293.15}{K}: modelling results obtained 
 with the electron collision cross sections from \cite{Song:2017} without 
 considering superelastic processes and with a modified set with and without 
 superelastic processes.}
 \label{fig: C2H2-mobSE}
\end{figure}

The influence of superelastic collisions is mostly visible in the drift velocity 
and mobility as shown in figure \ref{fig: C2H2-mobSE}. This figure compares the 
values of mobility and the longitudinal and transverse \textit{bulk} components 
of the diffusion tensor obtained with the original cross sections set with 
the results obtained using the modified set with and without the inclusion of 
superelastic processes. As predicted, the results of the modified set 
neglecting superelastic collisions are the same as those obtained with the 
original set. 
Superelastic collisions are responsible for a reduction of the electron mobility 
in the range of low reduced field, visible up to approximately \SI{20}{Td}. The 
influence on the components of the diffusion tensor is overall smaller than 
that on the mobility with the largest differences in the longitudinal component 
around \SI{10}{Td}.

 As the impact of superelastic collisions decreases remarkably above about 
\SI{20}{Td}, their influence on the effective ionization frequency and Townsend 
ionization coefficient is negligible.

\section{Concluding remarks}
\label{Sec 6}

We have investigated electron swarm parameters in \ce{C2H2}, \ce{C2H4} and 
\ce{C2H6} experimentally using a scanning drift tube, as well as computationally 
by solutions of the electron Boltzmann equation and via Monte Carlo simulation, 
corresponding to both time-of-flight and steady-state Townsend conditions. 
The measured data made it possible to derive the bulk drift velocity, the bulk 
longitudinal component of the diffusion tensor and the effective ionization frequency of
the electrons, for the wide range of the reduced electric field from 1 to \SI{1790}{Td}.
The measured TOF transport parameters as well as the effective SST ionization coefficient, 
deduced from the TOF swarm parameters, have been compared to experimental data obtained 
in previous studies. Here, generally 
good agreement with most of the transport parameters and the effective SST ionization 
coefficients obtained in these earlier 
studies was found. In the case of the drift velocity or the mobility, respectively, 
and the longitudinal component of the diffusion tensor we found disagreements at low or high 
values of $E/N$. 

The experimental data have undergone a correction procedure, which was supposed 
to quantify the errors caused by the dependence of the sensitivity of the 
detector of the drift cell on the energy distribution of the electrons in the 
swarm that may have a spatial dependence.

In particular, in case of \ce{C2H2} our measured drift velocities at low $E/N$ 
agree well with previous data of Bowman and Gordon~\cite{Bowman:1967} but not 
with the results of Cottrell and Walker~\cite{Cottrell:1965} as well as of
Nakamura \cite{Nakamura:2010}. 
Further measurements in this range are required to clarify this contradiction. 

The comparison of the experimental data was also carried out with swarm parameters 
resulting from various kinetic computations, which used the most 
recently recommended cross section sets~\cite{Song:2017,Fresnet:2002,Shishikura:1997}. 
Here, excellent agreement between electron Boltzmann equation  and MC simulation 
results verifies the computational approaches and data for the three gases. 
The agreement of the computed data with the present and previously measured values 
of the reduced effective ionization frequency and SST ionization 
coefficient was generally good. However, certain differences between kinetic 
computational and measured results found for the drift velocities and, especially, 
for the longitudinal component of the diffusion tensor illustrate the need for 
an improvement of the existing collision cross section sets for the three 
hydrocarbon gases considered.

We have also studied the influence of the thermally excited vibrational populations 
on the transport parameters. In the case of \ce{C2H2} we have found that this 
population has a significant value and superelastic collisions influence the drift 
velocity and the components of the diffusion tensor up to \SI{20}{Td}. The 
fitting of electron collision cross sections for this gas using swarm experiments 
should include these processes.

\ack

This work was partially supported by the Portuguese FCT--Fundação para a Ciência 
e a Tecnologia, under project UID/FIS/50010/2013, by the Hungarian Office for Research, 
Development and Innovation (NKFIH) grants K119357, K115805, by the  ÚNKP-19-3 New 
National Excellence Program of the Ministry for Innovation and Technology, and funded 
by the Deutsche Forschungsgemeinschaft (DFG, German Research Foundation) -- project number 327886311.
SD and DB are supported by Grants No. OI171037 
and III41011 from the Ministry of Education, Science and Technological Development 
of the Republic of Serbia. We thank Prof. Y. Nakamura for providing numerical values 
of measured electron transport parameters in \ce{C2H2} and  Mr T. Sz\H{u}cs for 
his contributions to the construction of the experimental apparatus.


\appendix

\section{Tables of experimental data}\label{App A}

\begin{table*}[ht]
  \caption{Measured bulk drift 
  	velocity $W$ of electrons in \ce{C2H2}, \ce{C2H4} 
  and \ce{C2H6} at \SI{293.15}{K}. The \textit{exptl.}
   values are the measured data
  while the \textit{cor.} ones are corrected for systematic effects ($E/N$ in \si{Td}, $W$ in \SI{e4}{m/s}).\\}
  \label{tab:WC2Hnwithneq246}
  \centering
  \sisetup{table-figures-integer=6, table-figures-decimal=2}
  \begin{tabular}{S[table-number-alignment=right]SS
                  S[table-number-alignment=right]SS
                  S[table-number-alignment=right]SS}
  \toprule
    {$E/N$} & \multicolumn{2}{c}{$W$} & {$E/N$} & \multicolumn{2}{c}{$W$} & {$E/N$} & \multicolumn{2}{c}{$W$} \\ 
  \cmidrule(r){2-3}\cmidrule(r){5-6}\cmidrule(r){8-9}
   & \textit{exptl.} & \textit{cor.} & & \textit{exptl.} & \textit{cor.} & & \textit{exptl.} & \textit{cor.} \\
  \midrule
  \multicolumn{9}{c}{\ce{C2H2}}\\ 
  \midrule
 3   &  1.43  & 1.43 & 125  & 8.36  & 8.33 &  450 & 29.6 & 30.3 \\
 5   &  2.36  & 2.35 & 145  & 9.38  & 9.35 &  480 & 31.9 & 32.7 \\
 9   &  3.48  & 3.47 & 160  & 10.1  & 10.2 &  540 & 36.0 & 37.1 \\
 21  &  4.88  & 4.86 & 170  & 10.8  & 10.9 &  600 & 40.4 & 41.7 \\                
 27  &  5.14  & 5.12 & 190  & 11.9  & 11.9 &  720 & 50.2 & 52.2 \\                
 34  &  5.38  & 5.36 & 220  & 13.6  & 13.7 &  830 & 58.4 & 61.0 \\                
 47  &  5.75  & 5.72 & 230  & 14.0  & 14.2 & 1043 & 73.5 & 77.8 \\                
 62  &  6.15  & 6.12 & 250  & 15.3  & 15.5 & 1120 & 77.9 & 82.8 \\                
 74  &  6.37  & 6.35 & 270  & 16.8  & 17.0 & 1330 & 93.7 & 108  \\                
 85  &  6.57  & 6.55 & 310  & 19.4  & 19.7 & 1530 & 108  & 117  \\ 
 96  &  7.17  & 7.15 & 360  & 23.0  & 23.4 & 1560 & 111  & 121  \\ 
 110 &  7.80  & 7.77 & 410  & 26.9  & 27.4 & 1790 & 122  & 135  \\ 
  \midrule
  \multicolumn{9}{c}{\ce{C2H4}}\\ 
  \midrule
1    &  2.33  & 2.27 & 50   & 5.69 & 5.53 & 310   & 19.8 & 19.4 \\
1.2  &  2.80  & 2.72 & 62   & 6.24 & 6.06 & 350   & 22.1 & 21.7 \\
1.5  &  2.89  & 2.81 & 69   & 6.60 & 6.41 & 410   & 25.6 & 25.3 \\
2    &  3.68  & 3.57 & 85   & 7.44 & 7.23 & 455   & 28.0 & 27.7 \\
2.5  &  3.99  & 3.88 & 100  & 8.51 & 8.27 & 480   & 29.7 & 29.4 \\                  
3.2  &  4.43  & 4.30 & 110  & 9.09 & 8.86 & 550   & 34.0 & 33.7 \\                  
4    &  4.74  & 4.60 & 125  & 10.2 & 9.96 & 600   & 37.0 & 36.7 \\                  
5    &  4.90  & 4.76 & 145  & 11.0 & 10.7 & 710   & 43.8 & 43.6 \\                  
7    &  5.02  & 4.88 & 160  & 11.8 & 11.5 & 840   & 50.9 & 50.9 \\                   
10   &  4.88  & 4.74 & 180  & 12.7 & 12.4 & 1040  & 64.7 & 65.2 \\
15   &  4.70  & 4.57 & 200  & 13.9 & 13.5 & 1140  & 70.5 & 71.2 \\
17   &  4.62  & 4.48 & 220  & 14.7 & 14.4 & 1330  & 82.6 & 83.9 \\
20   &  4.61  & 4.48 & 230  & 15.3 & 15.0 & 1510  & 82.1 & 84.0 \\ 
29   &  4.77  & 4.63 & 250  & 16.3 & 16.0 & 1570  & 86.1 & 88.3 \\
34   &  4.96  & 4.82 & 270  & 17.6 & 17.3 & 1760  & 93.6 & 96.6 \\
  \midrule
  \multicolumn{9}{c}{\ce{C2H6}}\\ 
  \midrule
1    &  3.23  &  3.14  &  48  & 5.36   & 5.22   & 350   & 20.3  & 20.1   \\
1.5  &  3.97  &  3.86  &  60  & 5.55   & 5.42   & 410   & 24.0  & 23.9   \\
2    &  4.58  &  4.45  &  70  & 5.68   & 5.55   & 470   & 27.8  & 27.7   \\
2.5  &  4.84  &  4.71  &  82  & 5.93   & 5.80   & 550   & 33.3  & 33.4   \\
3    &  5.02  &  4.88  &  100 & 6.46   & 6.32   & 610   & 37.3  & 37.6   \\
4    &  5.06  &  4.93  &  125 & 7.52   & 7.36   & 710   & 43.2  & 43.7   \\
5    &  5.31  &  5.17  &  145 & 8.44   & 8.28   & 715   & 42.4  & 43.0   \\
6    &  5.39  &  5.25  &  160 & 9.18   & 9.01   & 836   & 49.3  & 50.3   \\
7    &  5.49  &  5.34  &  180 & 10.2   & 10.0   & 900   & 53.8  & 55.0   \\
8.5  &  5.53  &  5.38  &  200 & 11.3   & 11.1   & 1002  & 60.3  & 62.0   \\
10   &  5.54  &  5.39  &  225 & 12.7   & 12.5   & 1170  & 71.8  & 74.5   \\
15   &  5.45  &  5.30  &  250 & 14.3   & 14.1   & 1398  & 77.0  & 80.9   \\
26   &  5.33  &  5.18  &  270 & 15.6   & 15.4   & 1606  & 82.9  & 88.1   \\
36   &  5.31  &  5.17  &  310 & 17.7   & 17.5   &       &       &        \\
  \bottomrule
  \end{tabular}
\end{table*}

\pagebreak 

\begin{table*}[ht]
  \caption{Measured longitudinal component of the diffusion tensor times gas number 
  density $D_{\rm L}N$ of electrons in \ce{C2H2}, \ce{C2H4} and \ce{C2H6} at 
  \SI{293.15}{K}. The \textit{exptl.} values are the measured data while the 
  \textit{cor.} ones are corrected for systematic effects ($E/N$ in \si{Td}, 
  $D_{\rm L}N$ in \SI{e24}{m^{-1}s^{-1}}).\\}
	\label{tab:NDLC2Hnwithneq246}
	\centering
  \sisetup{table-figures-integer=6, table-figures-decimal=2}
  \begin{tabular}{S[table-number-alignment=right]SS
                  S[table-number-alignment=right]SS
                  S[table-number-alignment=right]SS}
    \toprule
    {$E/N$} & \multicolumn{2}{c}{$D_\mathrm{L}N$} & 
    {$E/N$} & \multicolumn{2}{c}{$D_\mathrm{L}N$} & 
    {$E/N$} & \multicolumn{2}{c}{$D_\mathrm{L}N$} \\ 
    \cmidrule(r){2-3}\cmidrule(r){5-6}\cmidrule(r){8-9}
   & \textit{exptl.} & \textit{cor.} & & \textit{exptl.} & \textit{cor.} & & \textit{exptl.} & \textit{cor.} \\
		\midrule
		\multicolumn{9}{c}{\ce{C2H2}}\\ 
		\midrule
    21  & 0.84  & 0.53  & 160  & 1.10  & 1.02  & 480  & 2.68 & 3.00 \\ 
    27  & 0.63  & 0.42  & 170  & 1.17  & 1.10  & 540  & 2.82 & 3.24 \\ 
		34  & 0.48  & 0.34  & 190  & 1.31  & 1.28  & 600  & 3.03 & 3.75 \\ 
		47  & 0.51  & 0.38  & 220  & 1.44  & 1.43  & 720  & 3.46 & 4.28 \\ 
		62  & 0.47  & 0.37  & 230  & 1.53  & 1.53  & 830  & 3.70 & 4.78 \\ 
		74  & 0.46  & 0.38  & 250  & 1.67  & 1.68  & 1043 & 4.26 & 5.95 \\ 
		85  & 0.55  & 0.50  & 270  & 1.81  & 1.84  & 1120 & 4.36 & 6.25 \\ 
		96  & 0.67  & 0.57  & 310  & 1.97  & 2.04  & 1330 & 4.73 & 7.27 \\ 
		110 & 0.80  & 0.70  & 360  & 2.39  & 2.54  & 1530 & 4.98 & 8.14 \\ 
		125 & 0.83  & 0.74  & 410  & 2.39  & 2.59  & 1560 & 4.97 & 8.20 \\
		145 & 1.12  & 1.03  & 450  & 2.50  & 2.76  & 1790 & 5.35 & 9.43 \\
		\midrule
		\multicolumn{9}{c}{\ce{C2H4}}\\ 
		\midrule
		1   & 1.35  & 0.94  &  50   & 0.54   & 0.48   & 310  & 1.97  & 1.91  \\
		1.2 & 1.36  & 0.96  &  62   & 0.68   & 0.60   & 350  & 2.14  & 2.09  \\
		1.5 & 1.23  & 0.88  &  69   & 0.77   & 0.68   & 410  & 2.28  & 2.27  \\
		2   & 1.33  & 0.97  &  85   & 0.88   & 0.79   & 455  & 2.40  & 2.41  \\ 
		2.5 & 1.04  & 0.77  &  100  & 1.09   & 0.99   & 480  & 2.45  & 2.49  \\ 
		3.2 & 0.94  & 0.70  &  110  & 1.16   & 1.05   & 550  & 2.69  & 2.78  \\ 
		4   & 0.80  & 0.61  &  125  & 1.22   & 1.12   & 600   & 2.96  & 3.09 \\ 
		5   & 0.66  & 0.51  &  145  & 1.30   & 1.19   & 710   & 2.77  & 2.99 \\ 
		7   & 0.48  & 0.38  &  160  & 1.37   & 1.27   & 840   & 3.06  & 3.41 \\ 
		10  & 0.41  & 0.33  &  180  & 1.45   & 1.35   & 1040  & 3.64  & 4.26 \\  
		15  & 0.40  & 0.33  &  200  & 1.51   & 1.41   & 1140  & 3.99  & 4.77 \\  
		17  & 0.42  & 0.35  &  220  & 1.54   & 1.45   & 1330  & 4.49  & 5.61 \\  
    20  & 0.42  & 0.35  & 230   & 1.68   & 1.59   & 1510  & 3.33  & 4.32 \\ 
		29  & 0.42  & 0.36  & 250   & 1.69   & 1.60   & 1570  & 3.43  & 4.51 \\ 
		34  & 0.43  & 0.37  & 270   & 1.85   & 1.77   & 1760  & 3.70  & 5.06 \\ 
		\midrule
		\multicolumn{8}{c}{\ce{C2H6}}\\ 
		\midrule
	1   & 2.41  & 1.89  &  48  & 0.71   & 0.65   & 350   & 2.29   & 2.33   \\
	1.5 & 2.10  & 1.68  &  60  & 0.72   & 0.66   & 410   & 1.80   & 1.87   \\
	2   & 1.85  & 1.49  &  70  & 0.73   & 0.67   & 470   & 2.58   & 2.74   \\
	2.5 & 1.57  & 1.28  &  82  & 0.78   & 0.72   & 550   & 2.89   & 3.15   \\   
	3   & 1.42  & 1.17  &  100 & 0.94   & 0.87   & 610   & 3.31   & 3.67   \\  
	4   & 1.16  & 0.96  &  125 & 1.24   & 1.17   & 710   & 3.08   & 3.53   \\ 
	5   & 1.03  & 0.86  &  145 & 1.44   & 1.36   & 715   & 3.28   & 3.76   \\ 
	6   & 0.94  & 0.79  &  160 & 1.55   & 1.48   & 836   & 3.45   & 4.10   \\ 
	7   & 0.88  & 0.74  &  180 & 1.69   & 1.62   & 900   & 3.66   & 4.43   \\
	8.5 & 0.83  & 0.71  &  200 & 1.79   & 1.73   & 1002  & 3.73   & 4.65   \\
	10  & 0.79  & 0.68  &  225 & 1.90   & 1.85   & 1170  & 4.30   & 5.61   \\
  15  & 0.74  & 0.64  &  250 & 2.01   & 1.98   & 1398  & 4.75   & 6.58   \\ 
	26  & 0.73  & 0.65  &  270 & 2.09   & 2.07   & 1606  & 4.10   & 5.98   \\ 
	36  & 0.73  & 0.65  &  310 & 1.61   & 1.62   &       &        &        \\
	\bottomrule   
	\end{tabular}
\end{table*}

\pagebreak 

\begin{table*}[ht]
	\caption{Measured reduced effective ionization frequency   
		$\nu_\mathrm{eff}/N$ of electrons in \ce{C2H2},  \ce{C2H4} and  \ce{C2H6}
		at \SI{293.15}{K}. The \textit{exptl.} values are the measured data while the 
  \textit{cor.} ones are corrected for systematic effects ($E/N$ in \si{Td}, 
  $\nu_\mathrm{eff}/N$ in \SI{e-14}{m^3/s}).\\}
	\label{tab:nueffpNC2Hnwithneq246}
	\centering
  \sisetup{table-figures-integer=1, table-figures-decimal=0}
  \begin{tabular}{S[table-number-alignment=right]
                  S[table-column-width=20pt,table-number-alignment=left]
                  S[table-column-width=20pt,table-number-alignment=left]
                  S[table-number-alignment=right]
                  S[table-column-width=15pt,table-number-alignment=left]
                  S[table-column-width=15pt,table-number-alignment=left]
                  S[table-number-alignment=right]
                  S[table-column-width=10pt,table-number-alignment=left]
                  S[table-column-width=10pt,table-number-alignment=left]}
    \toprule
    {$E/N$} & \multicolumn{2}{c}{$\nu_\mathrm{eff}/N$} & 
    {$E/N$} & \multicolumn{2}{c}{$\nu_\mathrm{eff}/N$} & 
    {$E/N$} & \multicolumn{2}{c}{$\nu_\mathrm{eff}/N$} \\ 
    \cmidrule(r){2-3}\cmidrule(r){5-6}\cmidrule(r){8-9}
   & \textit{exptl.} & \textit{cor.} & & \textit{exptl.} & \textit{cor.} & & \textit{exptl.} & \textit{cor.} \\
		\midrule
		\multicolumn{9}{c}{\ce{C2H2}}\\ 
		\midrule
	145 & 0.00137  & 0.00136  & 310  & 0.0419  & 0.0422  & 830  & 0.853  & 0.879  \\
	160 & 0.00263  & 0.00264  & 360  & 0.0792  & 0.0800  & 1043 & 1.34   & 1.39   \\
	170 & 0.00325  & 0.00325  & 410  & 0.122   & 0.123   & 1120 & 1.50   & 1.56   \\
	190 & 0.00487  & 0.00488  & 450  & 0.158   & 0.161   & 1330 & 2.25   & 2.37   \\
	220 & 0.0109   & 0.0109   & 480  & 0.201   & 0.204   & 1530 & 3.06   & 3.25   \\
	230 & 0.0111   & 0.0111   & 540  & 0.263   & 0.268   & 1560 & 3.36   & 3.57   \\
	250 & 0.0171   & 0.0171   & 600  & 0.365   & 0.372   & 1790 & 3.80   & 4.07   \\
	270 & 0.0235   & 0.0237   & 720  & 0.603   & 0.619   &      &        &        \\ 
	\midrule
	\multicolumn{9}{c}{\ce{C2H4}}\\ 
	\midrule
	125 &  0.00106  &  0.00105  &  270 & 0.0241 & 0.0238 & 710   & 0.501 & 0.501 \\   
	145 &  0.00194  &  0.00191  &  310 & 0.0388 & 0.0385 & 840   & 0.696 & 0.698 \\  
	160 &  0.00310  &  0.00305  &  350 & 0.0592 & 0.0587 & 1040  & 1.20  & 1.21  \\  
	180 &  0.00463  &  0.00458  & 410  & 0.0881 & 0.0876 & 1140  & 1.36  & 1.37  \\  
	200 &  0.00682  &  0.00675  & 455  &  0.122 &  0.121 & 1330  & 1.83  & 1.85  \\    
	220 &  0.0101   &  0.00999  & 480  &  0.153 &  0.152 & 1510  & 2.30  & 2.33  \\    
	230 &  0.0138   &  0.0136   & 550  &  0.234 &  0.233 & 1570  & 2.54  & 2.59  \\    
	250 &  0.0186   &  0.0184   & 600  &  0.265 &  0.265 & 1760  & 2.90  & 2.97  \\    
	\midrule
	\multicolumn{9}{c}{\ce{C2H6}}\\ 
	\midrule
	100 & 0.000673 & 0.000675 & 270 & 0.0429 & 0.0430 & 715  & 0.571 & 0.581 \\ 
	125 & 0.00196  & 0.00197  & 310 & 0.0861 & 0.0863 & 836  & 0.776 & 0.792 \\ 
	145 & 0.00385  & 0.00386  & 350 & 0.0988 & 0.0991 & 900  & 0.929 & 0.949 \\
	160 & 0.00521  & 0.00522  & 410 & 0.221  & 0.222  & 1002 & 1.20  & 1.23  \\
	180 & 0.00667  & 0.00669  & 470 & 0.230  & 0.233  & 1170 & 1.61  & 1.65  \\ 
	200 & 0.0137   & 0.0137   & 550 & 0.356  & 0.361  & 1398 & 1.46  & 1.51  \\ 
	225 & 0.0214   & 0.0215   & 610 & 0.417  & 0.423  & 1606 & 1.82  & 1.89  \\
	250 & 0.0309   & 0.0310   & 710 & 0.657  & 0.669  &      &       &       \\
	\bottomrule
	\end{tabular}
\end{table*}

\pagebreak 

\begin{table*}[ht]
	\caption{Reduced effective SST ionization coefficient    
		$\alpha_\mathrm{eff}/N$ of electrons in \ce{C2H2},  \ce{C2H4} and  \ce{C2H6}
		at \SI{293.15}{K} calculated according to (\ref{eq:Blv1}) using the measured 
		values of $W$, $D_{\rm L}$ and $\nu_\mathrm{eff}$ ($E/N$ in \si{Td}, 
		$\alpha_\mathrm{eff}/N$ in \SI{e-20}{\square\metre}). The \textit{exptl.} values 
		are calculated with the measured data while the \textit{cor.} ones are 
		obtained with the values corrected for systematic effects. \\}
	\label{tab:alphaeffpNC2Hnwithneq246}
	\centering
  \sisetup{table-figures-integer=3, table-figures-decimal=0}
  \begin{tabular}{S[table-number-alignment=right]SS
                  S[table-number-alignment=right]SS
                  S[table-number-alignment=right]SS}
	\toprule
    {$E/N$} & \multicolumn{2}{c}{$\alpha_\mathrm{eff}/N$} & 
    {$E/N$} & \multicolumn{2}{c}{$\alpha_\mathrm{eff}/N$} & 
    {$E/N$} & \multicolumn{2}{c}{$\alpha_\mathrm{eff}/N$} \\ 
    \cmidrule(r){2-3}\cmidrule(r){5-6}\cmidrule(r){8-9}
   & \textit{exptl.} & \textit{cor.} & & \textit{exptl.} & \textit{cor.} & & \textit{exptl.} & \textit{cor.} \\
		\midrule
		\multicolumn{9}{c}{\ce{C2H2}}\\ 
		\midrule 
		145 & 0.0146  & 0.0145  & 310  & 0.221  & 0.219  &  830 & 1.63 & 1.65 \\    
		160 & 0.0260  & 0.0258  & 360  & 0.358  & 0.356  & 1043 & 2.08 & 2.15 \\    
		170 & 0.0301  & 0.0299  & 410  & 0.473  & 0.470  & 1120 & 2.19 & 2.28 \\ 
		190 & 0.0413  & 0.0411  & 450  & 0.561  & 0.558  & 1330 & 2.80 & 3.00 \\ 
		220 & 0.0809  & 0.0805  &  480 & 0.667  & 0.663  & 1530 & 3.37 & 3.74 \\
		230 & 0.0796  & 0.0791  &  540 & 0.778  & 0.775  & 1560 & 3.61 & 4.07 \\
		250  & 0.113   & 0.112  &  600 & 0.974  & 0.973  & 1790 & 3.71 & 4.33 \\               
		270  & 0.143   & 0.142  &  720 & 1.32   & 1.33   &      &      &      \\               
		\midrule
		\multicolumn{9}{c}{\ce{C2H4}}\\ 
		\midrule
		125   &  0.0104  &  0.0105  & 270  &  0.138  &  0.140  & 710  &  1.24 &  1.26 \\
		145   &  0.0177  &  0.0178  & 310  &  0.200  &  0.202  & 840  &  1.50 &  1.53 \\
		160   &  0.0264  &  0.0267  & 350  &  0.275  &  0.278  & 1040 &  2.10 &  2.16 \\ 
		180   &  0.0366  &  0.0370  & 410  &  0.355  &  0.358  & 1140 &  2.21 &  2.28 \\ 
		200   &  0.0495  &  0.0500  & 455  &  0.453  &  0.457  & 1330 &  2.57 &  2.69 \\ 
		220   &  0.0690  &  0.0698  & 480  &  0.539  &  0.544  & 1510 &  3.21 &  3.36 \\ 
		230   &  0.0905  &  0.0915  & 550  &  0.730  &  0.738  & 1570 &  3.41 &  3.59 \\                     
		250   &  0.115   &  0.117   & 600  &  0.765  &  0.772  & 1760 &  3.62 &  3.85 \\                     
		\midrule
		\multicolumn{9}{c}{\ce{C2H6}}\\ 
		\midrule
		100  &  0.0104  &  0.0107  & 270  & 0.286  & 0.291  & 715   & 1.52  & 1.57 \\
		125  &  0.0262  &  0.0268  & 310  & 0.511  & 0.519  & 836   & 1.80  & 1.86 \\
		145  &  0.0460  &  0.0470  & 350  & 0.518  & 0.524  & 900   & 2.00  & 2.07 \\
		160  &  0.0573  &  0.0585  & 410  & 0.994  & 1.01   & 1002  & 2.32  & 2.42 \\
		180  &  0.0662  &  0.0675  & 470  & 0.906  & 0.925  & 1170  & 2.68  & 2.82 \\   
		200  &  0.123   &  0.126   & 550  & 1.19   & 1.22   & 1398  & 2.20  & 2.29 \\   
		225  &  0.173   &  0.177   & 610  & 1.26   & 1.29   & 1606  & 2.50  & 2.60 \\  
		250  &  0.223   &  0.227   & 710  & 1.74   & 1.79   &       &       &      \\  
		\bottomrule
	\end{tabular}
\end{table*}

\clearpage 

\section{Statistical weights and statistical sums}\label{App B}

The fractional populations for the levels of a polyatomic molecule with $n_v$ modes
and vibrational quantum numbers $(v_1 v_2 v_3\ldots)$ are given by
\begin{equation}
\delta_{(v_1 v_2 v_3\ldots)} = \frac{g_{(v_1 v_2 v_3\ldots)}}{Q_v}
\exp\left(-\frac{\epsilon_{(v_1 v_2 v_3\ldots)}}{k_B T}\right)
\end{equation}
where $\epsilon_{(v_1 v_2 v_3\ldots)}$ is the level energy and $g$ the total 
statistical weight, 
\begin{equation}
 g_{(v_1 v_2 v_3\ldots)} = \prod_{n=1}^{n=n_v}\frac{(v_n+d_n-1)!}{v_n!(d_n-1)!}
\end{equation}
where $d_n$ is the degeneracy multiplicity for mode $n$, and $Q_v$ the vibrational 
statistical sum which, in the harmonic oscilator approximation for the vibrational
states, is
\begin{equation}
 Q_v = \prod_{n=1}^{n=n_v}(1-Z_n)^{-d_n}\,,\quad Z_n = \exp\{-hc\nu_n/k_B T\}
\end{equation}
where $\nu_n$ are the vibrational frequencies.

\section*{References}

\bibliography{C2HmBib}

\providecommand{\newblock}{}
\begin{thebibliography}{10}
\expandafter\ifx\csname url\endcsname\relax
  \def\url#1{{\tt #1}}\fi
\expandafter\ifx\csname urlprefix\endcsname\relax\def\urlprefix{URL }\fi
\providecommand{\eprint}[2][]{\url{#2}}

\bibitem{Adamovich:2014}
Adamovich I~V and Lempert W~R 2014 {\em Plasma Physics and Controlled Fusion\/}
  {\bf 57} 014001 doi: \url{10.1088/0741-3335/57/1/014001}

\bibitem{Starikovskiy:2013}
Starikovskiy A and Aleksandrov N 2013 {\em Progress in Energy and Combustion
  Science\/} {\bf 39} 61 -- 110 ISSN 0360-1285 doi:
  \url{10.1016/j.pecs.2012.05.003}

\bibitem{kosarev2009}
Kosarev I, Aleksandrov N, Kindysheva S, Starikovskaia l~S and Starikovskii A~Y
  2009 {\em Combustion and flame\/} {\bf 156} 221--233

\bibitem{kosarev2013}
Kosarev I, Pakhomov A, Kindysheva S, Anokhin E and Aleksandrov N 2013 {\em
  Plasma Sources Science and Technology\/} {\bf 22} 045018

\bibitem{kosarev2015}
Kosarev I, Kindysheva S, Aleksandrov N and Starikovskiy A~Y 2015 {\em
  Combustion and Flame\/} {\bf 162} 50--59

\bibitem{kosarev2016}
Kosarev I, Kindysheva S, Momot R, Plastinin E, Aleksandrov N and Starikovskiy
  A~Y 2016 {\em Combustion and Flame\/} {\bf 165} 259--271

\bibitem{Robertson:2002}
Robertson J 2002 {\em Materials Science and Engineering: R: Reports\/} {\bf 37}
  129 -- 281 ISSN 0927-796X doi: \url{10.1016/S0927-796X(02)00005-0}

\bibitem{Kumar:2010}
Kumar M and Ando Y 2010 {\em Journal of Nanoscience and Nanotechnology\/} {\bf
  10} 3739--3758 ISSN 1533-4880 doi: \url{10.1166/jnn.2010.2939}

\bibitem{Fonte:2010}
Fonte P and Peskov V 2010 {\em Plasma Sources Science and Technology\/} {\bf
  19} 034021 doi: \url{10.1088/0963-0252/19/3/034021}

\bibitem{VonKeudell:2001}
von Keudell A, Schwarz-Selinger T, Jacob W and Stevens A 2001 {\em Journal of
  Nuclear Materials\/} {\bf 290-293} 231 -- 237 ISSN 0022-3115 14th Int. Conf.
  on Plasma-Surface Interactions in Controlled Fusion Devices

\bibitem{Varanasi:1983}
Varanasi P, Giver L and Valero F 1983 {\em Journal of Quantitative Spectroscopy
  and Radiative Transfer\/} {\bf 30} 497 -- 504 ISSN 0022-4073 doi:
  \url{10.1016/0022-4073(83)90003-1}

\bibitem{Courtin:1984}
{Courtin} R, {Gautier} D, {Marten} A, {Bezard} B and {Hanel} R 1984 {\em
  Astrophysical Journal\/} {\bf 287} 899--916 doi: \url{10.1086/162748}

\bibitem{Hasegawa_Date:2015}
Hasegawa H and Date H 2015 {\em Journal of Applied Physics\/} {\bf 117} 133302
  doi: \url{10.1063/1.4916606}

\bibitem{Nakamura:2010}
Nakamura Y 2010 {\em Journal of Physics D: Applied Physics\/} {\bf 43} 365201
  doi: \url{10.1088/0022-3727/43/36/365201}

\bibitem{Cottrell:1968}
Cottrell T~L, Pollock W~J and Walker I~C 1968 {\em Trans. Faraday Soc.\/} {\bf
  64}(0) 2260--2266 doi: \url{10.1039/TF9686402260}

\bibitem{Bowman:1967}
Bowman C~R and Gordon D~E 1967 {\em The Journal of Chemical Physics\/} {\bf 46}
  1878--1883 doi: \url{10.1063/1.1840948}

\bibitem{Cottrell:1965}
Cottrell T~L and Walker I~C 1965 {\em Trans. Faraday Soc.\/} {\bf 61}(0)
  1585--1593 doi: \url{10.1039/TF9656101585}

\bibitem{Takatou:2011}
Takatou J, Sato H and Nakamura Y 2011 {\em Journal of Physics D: Applied
  Physics\/} {\bf 44} 315201 doi: \url{10.1088/0022-3727/44/31/315201}

\bibitem{Schmidt:1992}
Schmidt B and Roncossek M 1992 {\em Australian Journal of Physics\/} {\bf 45}
  351--364 doi: \url{10.1071/PH920351}

\bibitem{Wagner:1967}
Wagner E~B, Davis F~J and Hurst G~S 1967 {\em The Journal of Chemical
  Physics\/} {\bf 47} 3138--3147 doi: \url{10.1063/1.1712365}

\bibitem{Christophorou:1966}
Christophorou L~G, Hurst G~S and Hadjiantoniou A 1966 {\em The Journal of
  Chemical Physics\/} {\bf 44} 3506--3513 doi: \url{10.1063/1.1727257}

\bibitem{Hurst:1963}
Hurst G~S, O'Kelly L~B, Wagner E~B and Stockdale J~A 1963 {\em The Journal of
  Chemical Physics\/} {\bf 39} 1341--1345 doi: \url{10.1063/1.1734438}

\bibitem{Bortner:1957}
Bortner T~E, Hurst G~S and Stone W~G 1957 {\em Review of Scientific
  Instruments\/} {\bf 28} 103--108 doi: \url{10.1063/1.1715825}

\bibitem{Shishikura:1997}
Shishikura Y, Asano K and Nakamura Y 1997 {\em Journal of Physics D: Applied
  Physics\/} {\bf 30} 1610--1615 doi: \url{10.1088/0022-3727/30/11/010}

\bibitem{Kersten:1994}
Kersten H~J 1994 {\em {Messung der Driftgeschwindigkeit und des effektiven
  Townsendkoeffizienten von Elektronen bei hohen elektrischen Feldst\"arken}\/}
  Diploma thesis Ruprecht-Karls-Universit\"at Heidelberg

\bibitem{Heylen:1963}
Heylen A~E~D 1963 {\em The Journal of Chemical Physics\/} {\bf 38} 765--771
  doi: \url{10.1063/1.1733735}

\bibitem{Heylen:1978}
Heylen A~E~D 1978 {\em International Journal of Electronics\/} {\bf 44}
  367--374 doi: \url{10.1080/00207217808900831}

\bibitem{Watts:1979}
Watts M~P and Heylen A~E~D 1979 {\em Journal of Physics D: Applied Physics\/}
  {\bf 12} 695--702 doi: \url{10.1088/0022-3727/12/5/010}

\bibitem{Heylen:1975}
Heylen A~E~D 1975 {\em International Journal of Electronics\/} {\bf 39}
  653--660 doi: \url{10.1080/00207217508920532}

\bibitem{LeBlanc:1960}
LeBlanc O~H and Devins J~C 1960 {\em Nature\/} {\bf 188} 219--220 doi:
  \url{10.1038/188219a0}

\bibitem{Blevin:1984}
Blevin H~A and Fletcher J 1984 {\em Aust. J. Phys.\/} {\bf 37} 593--600 doi:
  \url{10.1071/PH840593}

\bibitem{Donko:2019}
Donk{\'{o}} Z, Hartmann P, Korolov I, Jeges V, Bo{\v{s}}njakovi{\'{c}} D and
  Dujko S 2019 {\em Plasma Sources Science and Technology\/} {\bf 28} 095007
  doi: \url{10.1088/1361-6595/ab3a58}

\bibitem{R}
Ramo S 1939 {\em Proc. IRE\/} {\bf 27} 584

\bibitem{Korolov:2016}
Korolov I, Vass M, Bastykova N~K and Donkó Z 2016 {\em Review of Scientific
  Instruments\/} {\bf 87} 063102 doi: \url{10.1063/1.4952747}

\bibitem{Korolov_2016}
Korolov I, Vass M and Donk{\'{o}} Z 2016 {\em Journal of Physics D: Applied
  Physics\/} {\bf 49} 415203 doi: \url{10.1088/0022-3727/49/41/415203}

\bibitem{Vass:2017}
Vass M, Korolov I, Loffhagen D, Pinh{\~{a}}o N and Donk{\'{o}} Z 2017 {\em
  Plasma Sources Science and Technology\/} {\bf 26} 065007 doi:
  \url{10.1088/1361-6595/aa6789}

\bibitem{S}
Shockley W 1938 {\em J. Appl. Phys.\/} {\bf 9} 635

\bibitem{SH}
Sirkis M and Holonyak N 1966 {\em Am. J. Phys.\/} {\bf 34} 943

\bibitem{Song:2017}
Song M~Y, Yoon J~S, Cho H, Karwasz G~P, Kokoouline V, Nakamura Y and Tennyson J
  2017 {\em Journal of Physical and Chemical Reference Data\/} {\bf 46} 013106
  doi: \url{10.1063/1.4976569}

\bibitem{Fresnet:2002}
Fresnet F, Pasquiers S, Postel C and Puech V 2002 {\em Journal of Physics D:
  Applied Physics\/} {\bf 35} 882--890 doi: \url{10.1088/0022-3727/35/9/308}

\bibitem{Leyh:1998}
Leyh H, Loffhagen D and Winkler R 1998 {\em Computer Physics Communications\/}
  {\bf 113} 33 -- 48 ISSN 0010-4655 doi: \url{10.1016/S0010-4655(98)00062-9}

\bibitem{Kumar:1980}
Kumar K, Skullerud H and Robson R 1980 {\em Aust. J. Phys.\/} {\bf 33} 343--448
  doi: \url{10.1071/PH800343b}

\bibitem{Segur:1983}
Segur P, Bordage M~C, Balaguer J~P and Yousfi M 1983 {\em J. Comput. Phys.\/}
  {\bf 50} 116–37

\bibitem{Dujko:2010}
Dujko S, White R~D, Petrovi\ifmmode~\acute{c}\else \'{c}\fi{} Z~L and Robson
  R~E 2010 {\em Phys. Rev. E\/} {\bf 81}(4) 046403 doi:
  \url{10.1103/PhysRevE.81.046403}

\bibitem{Dujko:2011}
Dujko S, White R~D, Petrovi{\'{c}} Z~L and Robson R~E 2011 {\em Plasma Sources
  Science and Technology\/} {\bf 20} 024013 doi:
  \url{10.1088/0963-0252/20/2/024013}

\bibitem{Kondo:1990}
Kondo K and Tagashira H 1990 {\em Journal of Physics D: Applied Physics\/} {\bf
  23} 1175--1183 doi: \url{10.1088/0022-3727/23/9/007}

\bibitem{Sh1972NBS}
Shimanouchi T 1972 Tables of {M}olecular {V}ibrational {F}requencies
  {C}onsolidated {V}olume {I} {Report NSDRS-NBS 39} National Bureau of
  Standards, Washington

\end{thebibliography}
\bibliographystyle{iopart-num}

\end{document}